\documentclass[12pt,a4paper]{iopart}

\usepackage{iopams}
\usepackage{setstack}
\usepackage{graphicx}
\usepackage{pict2e}
\usepackage{color}
\usepackage{hyperref}

\renewcommand{\P}{\mathbb{P}}

\renewcommand{\Im}{\mathrm{Im}}

\newcommand{\Z}{\mathbb{Z}}
\newcommand{\C}{\mathbb{C}}
\newcommand{\Ch}{\widehat{\mathbb{C}}}
\newcommand{\R}{\mathcal{R}}
\renewcommand{\[}{[\![}
\renewcommand{\]}{]\!]}
\newcommand{\p}{\mathfrak{p}}

\newcommand{\keywords}[1]{\noindent\textbf{Keywords:} #1}
\newcommand{\tfrac}[2]{\mbox{\small$\frac{#1}{#2}$}}
\renewcommand{\text}[1]{\mathrm{#1}}

\begin{document}

\title{From the Riemann surface of TASEP to ASEP}
\author{Sylvain Prolhac}
\address{Laboratoire de Physique Th\'eorique, IRSAMC, UPS, Universit\'e de Toulouse, France}


\begin{abstract}
We consider the asymmetric simple exclusion process (ASEP) with forward hopping rate $1$, backward hopping rate $q$ and periodic boundary conditions. We show that the Bethe equations of ASEP can be decoupled, at all order in perturbation in the variable $q$, by introducing a formal Laurent series mapping the Bethe roots of the totally asymmetric case $q=0$ (TASEP) to the Bethe roots of ASEP. The probability of the height for ASEP is then written as a single contour integral on the Riemann surface on which symmetric functions of TASEP Bethe roots live.\\


\end{abstract}

\begin{section}{Introduction}
The asymmetric simple exclusion process (ASEP) \cite{D1998.1,S2001.1,GM2006.1} is a Markov process featuring hard-core particles hopping on a one-dimensional lattice, at rate $1$ in the forward direction and rate $q$ in the backward direction, see figure~\ref{fig ASEP}. ASEP is a discrete model of KPZ universality \cite{KPZ1986.1,HHZ1995.1,KK2010.1,QS2015.1}, a prominent non-equilibrium setting with fluctuations exhibiting correlations on large scales, and for which several exact results have been observed in experiments \cite{TSSS2011.1}. The statistics of the ASEP height function for any $q\neq1$ converge in particular at large scales and under proper rescaling to that of the KPZ fixed point.

The literature on ASEP is large, especially concerning the system on an infinite line, for which exact expressions for the probability distribution of the height \cite{TW2009.1,SS2010.1,ACQ2011.1} have been obtained, building on earlier results \cite{S1997.1,PS2002.1,S2007.1} for the totally asymmetric model $q=0$ (TASEP). For the system in finite volume, with e.g. periodic boundary conditions, while some results are known about the spectral gap \cite{GS1992.1,K1995.1,P2017.2} and stationary large deviations \cite{DL1998.1,PM2009.1,P2010.1,S2011.1}, the complete evolution of the probability distribution of the height at finite times is only known so far for TASEP \cite{P2016.1,BL2018.1,BL2019.1}.

We consider in this paper ASEP with $N$ particles on $L$ sites and periodic boundary conditions. The deformed Markov matrix $M_{i}(q,\rme^{\gamma})$, with $\gamma$ a fugacity counting the current of particles between sites $i$ and $i+1$, generates the dynamics of the height function of ASEP. The matrix $M_{i}(q,\rme^{\gamma})$ is related by a similarity transformation to the Hamiltonian of a twisted XXZ quantum spin chain. ASEP is thus an integrable model, and Bethe ansatz dictates that the eigenstates of $M_{i}(q,\rme^{\gamma})$ may be expressed as linear combinations of plane waves. The $N$ momenta of the plane waves exhibit a rich algebraic structure, encoded in a system of $N$ coupled polynomial equations, the Bethe equations, which are independent of $i$ by translation invariance, and whose solutions are called the Bethe roots.

In the special case of TASEP, the Bethe equations have a mean field structure which can be solved in terms of a compact Riemann surface $\R$ \cite{P2020.2}, such that symmetric functions of $N$ Bethe roots are interpreted as meromorphic functions on $\R$. The genus of $\R$ goes to infinity in the scaling limit $L,N\to\infty$ to the KPZ fixed point, and the Riemann surface $\R$ then becomes \cite{P2020.1} the one on which half-integer polylogarithms are naturally defined. Similar results have also been obtained \cite{GP2020.1,GP2021.1} for TASEP with open boundaries. The Riemann surface approach for periodic TASEP allows to recover in a simpler way earlier results \cite{P2016.1,BL2018.1,BL2019.1} for height fluctuations of the KPZ fixed point with periodic boundaries.

The main appeal of the Riemann surface formalism for TASEP is that it allows to treat all eigenstates at once in a unified way. In particular, instead of dealing with individual eigenstates, which are complicated multiply valued functions of $\rme^{\gamma}$, it is much easier to deal with meromorphic functions living on a Riemann surface, due to the freedom that analyticity gives for integration contours.

We emphasize that the existence of a Riemann surface relating various eigenstates by analytic continuations is not a specific feature of TASEP, but is expected for generic interacting particle systems, for which a deformed Markov matrix $M(\rme^{\gamma})$ can be defined in a similar way as for TASEP. Indeed, the characteristic equation $\det(\lambda I-M(\rme^{\gamma}))=0$ verified by the eigenvalues $\lambda$ of $M(\rme^{\gamma})$ of any such process is a polynomial equation in both variables $\lambda$ and $\rme^{\gamma}$, and thus defines a Riemann surface. The specificity of TASEP is however that its Bethe ansatz structure allows for an elementary construction of this Riemann surface.

A natural question is thus whether the Riemann surface approach for TASEP can be extended to ASEP with arbitrary $q$, whose Bethe equations are much more intricate. This paper gives a partial answer, based on an all order perturbative solution of the Bethe equations around $q=0$. We introduce a function $Y_{\p,q}$ mapping the Bethe roots of TASEP to the Bethe roots of ASEP, where $Y_{\p,q}(y)$ is understood as a formal series in $q$ with coefficients meromorphic functions of $y\in\C$ and of $\p\in\R$. Imposing additional analyticity constraints, the Bethe equations are interpreted as the non-linear integro-differential equation (\ref{eq Y}) for $Y_{\p,q}$. Eigenstates of $M_{i}(q,g_{\p,q})$ for a specific choice of fugacity $g_{\p,q}$ are then written in terms of $Y_{\p,q}$. Our final result (\ref{P(Hi) stat}), (\ref{P(Hi) P0}) is an expression for the probability of the ASEP height function, written as a single contour integral of a formal series in $q$ depending on $Y_{\p,q}$.

The paper is organized as follows. In section~\ref{section Y}, we recall the solution of the TASEP Bethe equations in terms of the Riemann surface $\R$, and then construct the function $Y_{\p,q}$ solving the Bethe equations of ASEP. The probability of the height, written initially in terms of the ASEP Bethe roots, is then computed more explicitly in terms of $Y_{\p,q}$ in section~\ref{section proba}.

\begin{figure}
	\begin{center}
		\begin{picture}(70,60)(0,-2)
			\put(35,25){\circle{40}}
			\put(35,25){\circle{50}}
			\put(35,45){\line(0,1){5}}
			\put(42.6,43.48){\line(5,12){1.92}}
			\put(49.2,39.2){\line(1,1){3.54}}
			\put(53.48,32.6){\line(12,5){4.62}}
			\put(55,25){\line(1,0){5}}
			\put(53.48,17.4){\line(12,-5){4.62}}
			\put(49.2,10.8){\line(1,-1){3.54}}
			\put(42.6,6.52){\line(5,-12){1.92}}
			\put(35,5){\line(0,-1){5}}
			\put(27.4,6.52){\line(-5,-12){1.92}}
			\put(20.8,10.8){\line(-1,-1){3.54}}
			\put(16.52,17.4){\line(-12,-5){4.62}}
			\put(15,25){\line(-1,0){5}}
			\put(16.52,32.6){\line(-12,5){4.62}}
			\put(20.8,39.2){\line(-1,1){3.54}}
			\put(27.4,43.48){\line(-5,12){1.92}}
			\put(39.39,47.07){\circle*{2}}
			\put(53.71,37.50){\circle*{2}}
			\put(57.07,29.39){\circle*{2}}
			\put(39.39,02.93){\circle*{2}}
			\put(30.61,02.93){\circle*{2}}
			\put(22.50,06.29){\circle*{2}}
			\put(12.93,20.61){\circle*{2}}
			\put(22.50,43.71){\circle*{2}}
			\qbezier(62,28)(65,25)(62,22)
			\put(62,22){\vector(-1,-1){0.5}}
			\put(64.5,23){$1$}
			\qbezier(18.0294, 3.7868)(13.7868,3.7868)(13.7868,8.02944)
			\put(13.7868,8.02944){\vector(0,1){0.5}}
			\put(12.5,1.5){$1$}
			\qbezier(8,22)(5,25)(8,28)
			\put(8,28){\vector(1,1){0.5}}
			\put(3.5,23.5){$1$}
			\qbezier(21.8959,48.7967)(23.5195,52.7164)(27.4392,51.0928)
			\put(27.4392, 51.0928){\vector(1.30656,-0.541196){0.5}}
			\put(23.5,53){$1$}
			\qbezier(42.5608,51.0928)(46.4805,52.7164)(48.1041,48.7967)
			\put(48.1041,48.7967){\vector(0.541196,-1.30656){0.5}}
			\put(46,52.5){$1$}
			\qbezier(56.2132,41.9706)(56.2132,46.2132)(51.9706,46.2132)
			\put(51.9706, 46.2132){\vector(-1,0){0.5}}
			\put(56,46.5){$q$}
			\qbezier(42.5608,-1.0928)(46.4805,-2.71639)(48.1041,1.2033)
			\put(48.1041,1.2033){\vector(0.382683,0.92388){0.5}}
			\put(46,-4){$q$}
			\qbezier(8.9072,17.4392)(7.28361,13.5195)(11.2033,11.8959)
			\put(11.2033,11.8959){\vector(0.92388,-0.382683){0.5}}
			\put(5.5,13){$q$}
			\qbezier(18.0294,46.2132)(13.7868,46.2132)(13.7868,41.9706)
			\put(13.7868,41.9706){\vector(0,-1){0.5}}
			\put(13,47){$q$}
			\qbezier(38,52)(35,55)(32,52)
			\put(32,52){\vector(-1,-1){0.5}}
			\put(34.5,56){$q$}
		\end{picture}
	\end{center}
	\caption{Hopping rates of all allowed movements of particles for ASEP in a generic configuration with $N=8$ particles on a periodic lattice with $L=16$ sites.}
	\label{fig ASEP}
\end{figure}
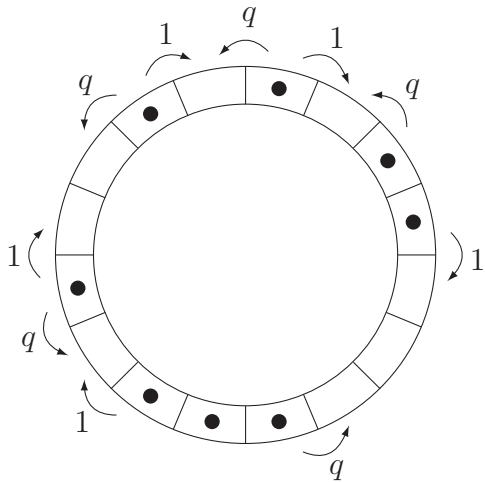
\end{section}

\begin{section}{Integro-differential equation for the function \texorpdfstring{$Y_{\p,q}$}{Y\_pq}}
\label{section Y}
In this section, we first recall some results about the Bethe equations for TASEP and their solution in terms of a Riemann surface $\R$, then consider the Bethe equations of ASEP and solve them perturbatively in $q$ by introducing a function $Y_{\p,q}$ mapping Bethe roots of TASEP to Bethe roots of ASEP.

\begin{subsection}{TASEP Bethe equations and Riemann surface \texorpdfstring{$\R$}{R}}
\label{section TASEP}
In this section, we recall some results from \cite{P2020.2} about the compact Riemann surface $\R$ (called $\R_{N}$ in \cite{P2020.2}) in terms of which Bethe ansatz for TASEP can be conveniently formulated.

Any eigenstate of the locally deformed Markov matrix $M_{i}(0,g)$ of TASEP may be expressed in terms of $N$ Bethe roots $y_{j}$ solution of the Bethe equations $g\,(1-y_{j})^{L}=(-1)^{N-1}\prod_{k}\frac{y_{j}}{y_{k}}$. The Bethe equations of TASEP have a kind of mean field nature, since any $y_{j}$ is coupled to the other $y_{k}$ only through their product. Introducing \footnote{The parameter $B$ defined here is equal to $B=N^{N}(L-N)^{L-N}C/L^{L}$ in the notations of \cite{P2020.2}.} the parameter $B=g\prod_{k}y_{k}$, the Bethe roots are then solutions of a polynomial equation of degree $L$, $P(y_{j},B)=0$, with
\begin{equation}
\label{P(y,B)}
P(y,B)=B(1-y)^{L}+(-1)^{N}y^{N}\;.
\end{equation}
The set of all ${L}\choose{N}$ eigenstates for a fixed value of $B$ correspond to all possible ways to choose $N$ roots $y$ of $P(y,B)=0$ among $L$.

Following \cite{P2020.2}, from the polynomial equation $P(y,B)=0$ one can define $L$ Bethe root functions $y_{j}(B)$, $j=1,\ldots,L$, analytic for $B\in\C\setminus\mathbb{R}^{-}$, with possible branch points $0$, $B_{*}$ and $\infty$, where
\begin{equation}
\label{B*}
B_{*}=-\frac{N^{N}(L-N)^{L-N}}{L^{L}}\;.
\end{equation}
We choose the same labelling of the functions $y_{j}$ as in \cite{P2020.2}, see figure~\ref{fig yj TASEP}, which is such that
\begin{equation}
\label{yj(infinity)}
1-y_{j}(B)\underset{|B|\to\infty}{\simeq}\rme^{-\frac{2\rmi\pi}{L}\big(j-\frac{N+1}{2}\big)}B^{-1/L}\;,
\end{equation}
where $B^{-1/L}$ is defined with the usual branch cut $B\in\mathbb{R}^{-}$. With this choice, one has furthermore
\begin{equation}
\label{yj(0)}
y_{j}(B)\underset{|B|\to0}{\simeq}\Bigg\{
\begin{array}{lcr}
\rme^{\frac{2\rmi\pi}{N}\big(j-\frac{N+1}{2}\big)}B^{1/N} && 1\leq j\leq N\\
\rme^{-\frac{2\rmi\pi}{L-N}\big(j-\frac{L+N+1}{2}\big)}B^{-1/(L-N)} && N+1\leq j\leq L
\end{array}
\end{equation}
and
\begin{equation}
\label{yj(B*)}
\fl\hspace{10mm}
N+(L-N)y_{j}(B)\underset{B\to B_{*}}{\sim}\left\{
\begin{array}{lcc}
\sqrt{B-B_{*}} && j\in\{1,N+1\}\;\;\text{and}\;\;\Im\,B<0\\
\sqrt{B-B_{*}} && j\in\{N,L\}\;\;\text{and}\;\;\Im\,B>0\\
(B-B_{*})^{0} && \text{otherwise}
\end{array}
\right.\;,
\end{equation}
with the same branch cut $\mathbb{R}^{-}$ for the fractional powers. Analytic continuations across the branch cut $(-\infty,B_{*})$ generates cyclic permutation of all $y_{j}$, while analytic continuations across the branch cut $(B_{*},0)$ generates independent cyclic permutation of the $y_{j}$ with $1\leq j\leq N$ and with $N+1\leq j\leq L$, see figure~\ref{fig yj TASEP}.

\begin{figure}
	\begin{center}
	\begin{tabular}{l}
		\includegraphics[width=150mm]{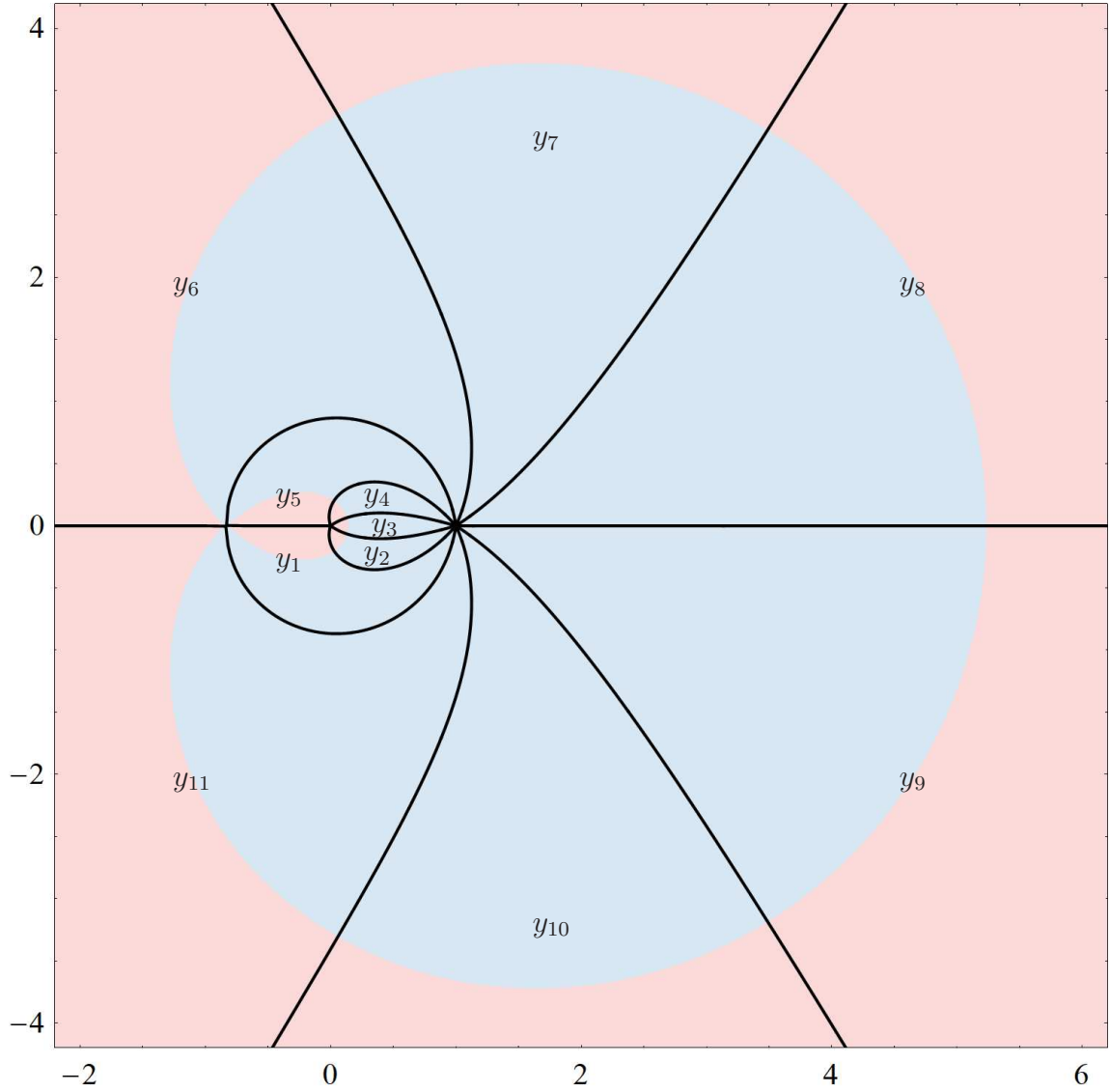}
		\begin{picture}(0,0)
			\put(-115,72){$y_{1}$}
			\put(-103,73){$y_{2}$}
			\put(-102,76.9){$y_{3}$}
			\put(-103,81){$y_{4}$}
			\put(-115,81){$y_{5}$}
			\put(-129,110){$y_{6}$}
			\put(-80,130){$y_{7}$}
			\put(-30,110){$y_{8}$}
			\put(-30,42){$y_{9}$}
			\put(-80,22){$y_{10}$}
			\put(-129,42){$y_{11}$}
		\end{picture}
	\end{tabular}
	\end{center}
	\caption{Graphical representation of the TASEP Bethe root functions $y_{j}(B)$ for a system with $L=11$, $N=5$. Each domain delimited by black curves represents the image by the corresponding function $y_{j}$ of the complex plane minus the cut $\mathbb{R}^{-}$. The red area (outer area plus the small inclusion within the inner area) corresponds to $|B|<|B_{*}|$ while the blue, inner area corresponds to $|B|>|B_{*}|$.}
	\label{fig yj TASEP}
\end{figure}

Symmetric functions of $N$ distinct Bethe roots then live on a Riemann surface $\R$, built by gluing together sheets $\C_{J}$ labelled by sets $J\subset\[1,L\]$ of $N$ integers according to analytic continuations of the functions $y_{j}(B)$ across the cuts $(-\infty,B_{*})$ and $(B_{*},0)$. The points $\p\in\R$ may then be labelled as $\p=[B,J]$, $B\in\Ch$ (where $\Ch=\C\cup\{\infty\}$ is the Riemann sphere), $J\subset\[1,L\]$ of cardinal $|J|=N$, with some identifications at the branch points $B\in\{0,B_{*},\infty\}$.

The Riemann surface $\R$ has several connected components if $L$ and $N$ are not co-prime. The total genus of $\R$, which was computed in \cite{P2020.2} for small values of $L$ and $N$, is strictly positive if $3\leq N\leq L-3$, $(N,L)\neq(3,6)$, and is conjectured to go to infinity exponentially fast when $L,N\to\infty$ with fixed density of particles $N/L$.

To any point $\p=[B,J]\in\R$ with $B\notin\{0,B_{*},\infty\}$ corresponds a single eigenstate of $M_{i}(0,g)$ for $g=g_{\p,0}$, with
\begin{equation}
\label{g0}
g_{[B,J],0}=\frac{B}{\Pi_{[B,J]}}\;,
\end{equation}
and
\begin{equation}
\label{Pi}
\Pi_{[B,J]}=\prod_{j\in J}y_{j}(B)\;.
\end{equation}
The corresponding eigenvalue is in particular equal to $\eta_{\p}$, with
\begin{equation}
\label{eta}
\eta_{[B,J]}=\sum_{j\in J}\frac{y_{j}(B)}{1-y_{j}(B)}\;.
\end{equation}
Exact expressions are known for various scalar products of eigenvectors, see \cite{P2020.2}.

\begin{table}
	\begin{center}
		\begin{tabular}{|c|c|c|}\hline
			Function & poles & zeroes\\\hline&&\\
			$\Pi_{\p}$ & $[0,J]$ & $[0,J]$\\&&\\\hline&&\\
			$\eta_{\p}$ & $[\infty,J]$ & \ldots\\&&\\\hline&&\\
			$\bar{\Pi}_{\p}$ & $[0,J]$ & $[\infty,J]$\\&&\\\hline&&\\
			$\Pi^{*}_{\p}$ & $[0,J]$ & \begin{tabular}{c}$[B_{*}+\rmi0^{+},J]$ \,with\, $\{N,L\}\cap J\neq\emptyset$\\$[B_{*}-\rmi0^{+},J]$ \,with\, $\{1,N+1\}\cap J\neq\emptyset$\end{tabular}\\&&\\\hline&&\\
			$V^{2}_{\p}$ & $[0,J]$ & \begin{tabular}{c}$[0,J]$\\$[\infty,J]$\\$[B_{*}+\rmi0^{+},J]$ \,with\, $\{N,L\}\subset J$\\$[B_{*}-\rmi0^{+},J]$ \,with\, $\{1,N+1\}\subset J$\end{tabular}\\&&\\\hline&&\\
			$\mu_{\p}$ & $[B_{*}\pm\rmi0^{+},J]$ & \ldots\\&&\\\hline&&\\
			\begin{tabular}{c}$\alpha_{\p,m}$\\[2mm]$m\in\Z^{*}$\end{tabular} & $[0,J]$ & \ldots\\&&\\\hline
		\end{tabular}
	\end{center}
	\caption{Potential location of the poles and zeroes of some meromorphic functions of the variable $\p\in\R$ defined in section~\ref{section TASEP}. Additional constraints on the set $J$ are needed in some cases for $\p$ to be a pole or a zero. The entries of the table marked as \ldots correspond to zeroes of sums, which are not located at ``simple'' points $[B,J]$ with $B\in\{0,B_{*},\infty\}$.}
	\label{table poles zeroes}
\end{table}

Anticipating the next sections, we finally define
\begin{eqnarray}
\label{Pib}
&& \bar{\Pi}_{[B,J]}=\prod_{j\in J}(1-y_{j}(B))\\
\label{Pi*}
&& \Pi^{*}_{[B,J]}=\prod_{j\in J}(N+(L-N)y_{j}(B))\\
\label{V2}
&& V^{2}_{[B,J]}=\prod_{j,k\in J\atop j<k}(y_{j}(B)-y_{k}(B))^{2}\\
\label{mu}
&& \mu_{[B,J]}=\sum_{j\in J}\frac{y_{j}(B)}{N+(L-N)y_{j}(B)}\\
\label{alpham}
&& \alpha_{[B,J],m}=\sum_{j\in J}\frac{1}{y_{j}(B)^{m}}\;,
\end{eqnarray}
with $m\in\mathbb{Z}$ for $\alpha_{\p,m}$ (the definition of $\mu_{\p}$ differs from $\mu_{1}(\p)$ in \cite{P2020.2} by a factor $-\frac{N}{L-N}$).

The coefficients defined above are meromorphic functions of $\p=[B,J]\in\R$, whose location of poles and zeroes, given in table~\ref{table poles zeroes}, follows from (\ref{yj(infinity)})-(\ref{yj(B*)}). A key point for height fluctuations in section~\ref{section proba}, proved in details in \cite{P2020.2}, is that while the function $\p\mapsto V^{2}_{\p}/\Pi^{*}_{\p}$ has simple poles at the points of the form $\p=[B_{*},J]$, the differential $\rmd B\,V^{2}_{[B,J]}/\Pi^{*}_{[B,J]}$ is however holomorphic at these points after going to proper local coordinates on $\R$.
\end{subsection}

\begin{subsection}{Strategy for the solution of the ASEP Bethe equations}
\label{section strategy}
Any eigenstate of the locally deformed Markov matrix $M_{i}(q,g)$ of ASEP with arbitrary asymmetry $q$ is given in terms of $N$ Bethe roots $Y_{j}$ solution of the Bethe equations
\begin{equation}
\label{Bethe eq ASEP}
g\,\Big(\frac{1-Y_{j}}{1-q\,Y_{j}}\Big)^{L}=-\prod_{k}\frac{Y_{j}-q\,Y_{k}}{q\,Y_{j}-Y_{k}}\;,
\end{equation}
see e.g. \cite{GS1992.2,PM2008.1}. Unlike in the TASEP case $q=0$, the Bethe roots $Y_{j}$ are now coupled to each other in a non-trivial way, and it is no longer possible to decouple them by introducing a single parameter $B$. Instead, the strategy developed in this paper to ``solve'' the Bethe equations of ASEP (\ref{Bethe eq ASEP}) is based on a systematic perturbative approach, consisting in expanding the Bethe root $Y_{j}$ as a power series in $q$, with coefficients depending on a point $\p$ of the Riemann surface $\R$ and on the TASEP Bethe root $y_{j}$.

As explained in the previous section, to each set of $N$ integers $J\subset\[1,L\]$ is associated an eigenstate of the locally deformed Markov matrix $M_{i}(0,g)$ of TASEP. This eigenstate is characterized by the Bethe root functions $\{y_{j}(B),j\in J\}$, and corresponds to the sheet $\C_{J}$ of the Riemann surface $\R$. The fugacity $g$ is related to the parameter $B$ by $g=g_{[B,J],0}$, with $g_{\p,0}$ the meromorphic function of $\p\in\R$ defined in (\ref{g0}).

By continuity, we may label the eigenstates of ASEP by the same sets $J$, at least for small enough $q$, and away from the branch points $B\in\{0,B_{*},\infty\}$ where several eigenstates coalesce. For a given point $\p=[B,J]\in\R$, we thus label the Bethe roots solution of the ASEP Bethe equations (\ref{Bethe eq ASEP}) as $Y_{j}=Y_{\p,q,j}$, $j\in J$, such that $Y_{j}\to y_{j}(B)$ when $q\to0$. Then, the eigenstates of the locally deformed Markov matrix $M_{i}(q,g)$ of ASEP can be obtained from the Bethe roots $Y_{j}$, for a fugacity $g=g_{\p,q}$ converging to $g_{\p,0}$ when $q\to0$.

Introducing the TASEP solution, which verifies $P(y_{j}(B),B)=0$ with $P$ defined in (\ref{P(y,B)}), the Bethe equations (\ref{Bethe eq ASEP}) can then be rewritten with the notations above as
\begin{equation}
\label{Bethe eq ASEP/TASEP}
\frac{g_{\p,q}}{B}\,\frac{y_{j}(B)^{N}}{(1-y_{j}(B))^{L}}\,\Big(\frac{1-Y_{\p,q,j}}{1-q\,Y_{\p,q,j}}\Big)^{L}=\prod_{k\in J}\frac{Y_{\p,q,j}-q\,Y_{\p,q,k}}{Y_{\p,q,k}-q\,Y_{\p,q,j}}\;,
\end{equation}
with $\p=[B,J]$. We now make the key assumption that there exists a function $Y_{\p,q}$, meromorphic at each order in $q$, such that $Y_{\p,q,j}=Y_{\p,q}(y_{j}(B))$, and for which (\ref{Bethe eq ASEP/TASEP}) holds even after replacing $y_{j}(B)$ by an arbitrary variable $y$, i.e.
\begin{equation}
\label{Bethe eq Y}
\frac{g_{\p,q}}{B}\,\frac{y^{N}}{(1-y)^{L}}\,\Big(\frac{1-Y_{\p,q}(y)}{1-q\,Y_{\p,q}(y)}\Big)^{L}=\prod_{k\in J}\frac{Y_{\p,q}(y)-q\,Y_{\p,q}(y_{k}(B))}{Y_{\p,q}(y_{k}(B))-q\,Y_{\p,q}(y)}
\end{equation}
with $\p=[B,J]$. The Bethe equation (\ref{Bethe eq Y}) is rewritten below as a non-linear integro-differential equation (\ref{eq Y}).

The whole procedure depends crucially on the choice of $g_{[B,J],q}$ relating the fugacity and the variable $B$. We choose in the following $g_{\p,q}$ so that, at each order in perturbation near $q=0$, the function $Y_{\p,m}(y)$ is a Laurent polynomial in $y$, which constrains uniquely $g_{\p,q}$. We observe in section~\ref{section P(H)} that this particular choice of $g_{\p,q}$ also makes the pole structure on the Riemann surface $\R$ of the integrand in the expression for the probability of the height of ASEP identical to the pole structure found in \cite{P2020.2} for TASEP, and thus allows greater freedom in the choice of contours of integration.
\end{subsection}

\begin{subsection}{Perturbative solution in \texorpdfstring{$q$}{q}}
We introduce the small $q$ expansions
\begin{eqnarray}
\label{gq[hm]}
&& g_{\p,q}=\frac{B}{\Pi_{\p}}\Big(1+\sum_{m=1}^{\infty}h_{\p,m}\,q^{m}\Big)\\
\label{Y[Wm]}
&& Y_{\p,q}(y)=y\Big(1+\sum_{m=1}^{\infty}W_{\p,m}(y)\,q^{m}\Big)\;,
\end{eqnarray}
with $\p=[B,J]$, and $\Pi_{\p}$ defined in (\ref{Pi}) a meromorphic function of the variable $\p\in\R$.

\begin{subsubsection}{First order in $q$}\hfill\\
\label{section order q^1}
At first order in $q$, inserting the expansions (\ref{gq[hm]}), (\ref{Y[Wm]}) into (\ref{Bethe eq Y}) implies
\begin{equation}
\label{Bethe eq Y order 1}
\fl\hspace{5mm} (L-\alpha_{\p,1})\,y+\frac{\alpha_{\p,-1}}{y}+h_{\p,1}-W_{\p,1}(y)\,\frac{N+(L-N)y}{1-y}+\sum_{k\in J}W_{\p,1}(y_{k}(B))=0\;,
\end{equation}
with $\p=[B,J]$ and $\alpha_{\p,m}$, $m\in\mathbb{Z}$ the meromorphic functions of $\p\in\R$ defined in (\ref{alpham}).

As explained in section~\ref{section strategy}, we look for an expansion such that the functions $W_{\p,m}(y)$ are Laurent polynomials in $y$. In particular, $W_{\p,1}(y)=\sum_{r=r_{0}}^{r_{1}}\frac{w_{\p,r}}{y^{r}}$ for some $r_{0}<r_{1}\in\mathbb{Z}$. Equation (\ref{Bethe eq Y order 1}) then rewrites
\begin{equation}
(L-\alpha_{\p,1})\,y+\frac{\alpha_{\p,-1}}{y}+h_{\p,1}+\sum_{r=r_{0}}^{r_{1}}w_{\p,r}\Big(\alpha_{\p,r}-\frac{N+(L-N)y}{y^{r}(1-y)}\Big)=0\;.
\end{equation}
The limits $y\to0$ and $y\to\infty$ imply respectively $r_{1}=1$ and $r_{0}=-1$. Using $\alpha_{\p,0}=N$, we finally obtain
\begin{equation}
h_{\p,1}=\frac{L\,\alpha_{\p,-1}(L-\alpha_{\p,1})}{N(L-N)}
\end{equation}
and
\begin{equation}
W_{\p,1}(y)=(1-y)\Big(\frac{\alpha_{\p,-1}}{Ny}+\frac{L-\alpha_{\p,1}}{L-N}\Big)\;.
\end{equation}
We observe that $W_{\p,1}(1)=0$, and that the expansion $g_{\p,q}\simeq g_{\p,0}(1+q\,h_{\p,1})$ is indeed fully constrained by the requirement that the coefficients in the $q\to0$ expansion of $Y_{\p,q}$ are Laurent polynomials.
\end{subsubsection}

\begin{subsubsection}{Higher orders in $q$}\hfill\\
\label{section order q^m}
We want to show that at each order in $q$, there exists a unique choice for the coefficient $h_{\p,m}$ such that the functions $W_{\p,m}$ are Laurent polynomials. The coefficients of these Laurent polynomials are furthermore (ordinary) polynomials in the $\alpha_{\p,n}$, $n\in\mathbb{Z}^{*}$ with coefficients rational functions of $L$ and $N$, and one has the identity $W_{\p,m}(1)=0$.

We proceed by recursion on the order in $q$. From the results in the previous section, the properties above are true at first order in $q$, and we assume that it is also true up to order $q^{m-1}$. Inserting the expansions (\ref{gq[hm]}), (\ref{Y[Wm]}) into the Bethe equation (\ref{Bethe eq Y}), we obtain at order $q^{m}$ in terms of $\p=[B,J]$
\begin{equation}
\label{Bethe eq Y order m}
h_{\p,m}-\frac{N+(L-N)y}{1-y}\,W_{\p,m}(y)+\sum_{k\in J}W_{\p,m}(y_{k}(B))=\ldots\;,
\end{equation}
where the right side is a Laurent polynomial in $y$ with coefficients polynomials in the $\alpha_{[B,J],n}$, themselves with coefficients rational functions of $L$ and $N$.

We now assume that $W_{\p,m}(y)$ is a Laurent polynomial in $y$. The sum over $k\in J$ in (\ref{Bethe eq Y order m}) then gives further coefficients $\alpha_{\p,n}$. Cancellation of the pole at $y=1$ then gives $W_{\p,m}(1)=0$, which constrains the constant term of the Laurent polynomial $W_{\p,m}(y)$ in terms of its other coefficients. The non-constant terms in the partial fraction decomposition in $y$ of $\frac{N+(L-N)y}{1-y}\,W_{\p,m}(y)$ then must match the right side of (\ref{Bethe eq Y order m}), which fixes both the degrees of the Laurent polynomial $W_{\p,m}(y)$ and its remaining coefficients. Finally, the constant terms in $y$ of (\ref{Bethe eq Y order m}) constrains the coefficient $h_{\p,m}$ in a unique way.
\end{subsubsection}
\end{subsection}

\begin{subsection}{Sums over \texorpdfstring{$k\in J$}{k in J} as contour integrals}
In order to reformulate more concisely the perturbative solution of the Bethe equations describe above, a better way to deal with sums over $k\in J$ of functions of $y_{k}(B)$ is needed.

For any series $f_{q}(y)=\sum_{m=0}^{\infty}c_{m}(y)q^{m}$ with coefficients $c_{m}(y)$ Laurent polynomials in $y$, one can write $f_{q}(y)=\sum_{r\in\mathbb{Z}}\frac{d_{q,r}}{y^{r}}$, where only a finite number of coefficients $d_{q,r}$ contribute at any order in $q$. Then, one has from (\ref{alpham})
\begin{equation}
\sum_{k\in J}f_{q}(y_{k}(B))=\sum_{r\in\mathbb{Z}}\sum_{k\in J}\frac{d_{q,r}}{y_{k}(B)^{r}}=\sum_{r\in\mathbb{Z}}\alpha_{\p,r}\,d_{q,r}\;,
\end{equation}
where the last expression has only a finite number of terms at each order in $q$. Introducing the Laurent series
\begin{equation}
A_{\p}(z)=\sum_{r=-\infty}^{\infty}\alpha_{\p,r}z^{r}\;,
\end{equation}
we obtain
\begin{equation}
\label{sum[int] regular}
\sum_{k\in J}f_{q}(y_{k}(B))=\oint\frac{\rmd z}{2\rmi\pi z}\,A_{[B,J]}(z)f_{q}(z)\;,
\end{equation}
where the integration is over a counter-clockwise contour encircling $0$ once.

We note that $A_{\p}(z)$ is only a \emph{formal} Laurent series in $z$. Indeed, from (\ref{alpham}), one has $A_{[B,J]}(z)=\sum_{k\in J}\sum_{r\in\mathbb{Z}}(z/y_{k}(B))^{r}$, where the sum over $r$ is divergent for any $z\in\C$. This is however not a problem in (\ref{sum[int] regular}) since at any order in $q$, $f_{q}(z)$ is a well defined Laurent \emph{polynomial} in $z$, and the residue at $z=0$ of the product $A_{\p}(z)f_{q}(z)$ makes sense, at least as a formal power series in $q$.

The summation formula (\ref{sum[int] regular}) can alternatively be derived without considering the Laurent expansion of $f_{q}$. We write $A_{\p}(z)=A_{\p}^{+}(z)+A_{\p}^{-}(z)$ with $A_{\p}^{+}(z)=\sum_{r=1}^{\infty}\alpha_{\p,r}z^{r}$ and $A_{\p}^{-}(z)=\sum_{r=-\infty}^{0}\alpha_{\p,r}z^{r}$. From the definition (\ref{alpham}) of the $\alpha_{\p,r}$, both $A_{\p}^{+}(z)$ and $A_{\p}^{-}(z)$ can in fact be expressed, in distinct domains for the variable $z$, in terms of the resolvant of the TASEP Bethe roots: one has $A_{[B,J]}^{+}(z)=-\sum_{k\in J}\frac{z}{z-y_{k}(B)}$ for any $z$ in the disk $|z|<\min\limits_{k\in J}y_{k}(B)$, and $A_{[B,J]}^{-}(z)=\sum_{k\in J}\frac{z}{z-y_{k}(B)}$ for any $z$ in the annulus $|z|>\max\limits_{k\in J}y_{k}(B)$. Thus, $\oint\frac{\rmd z}{2\rmi\pi z}\,A_{\p}(z)f_{q}(z)=
-\oint_{|z|<\min\limits_{k\in J}y_{k}(B)}\frac{\rmd z}{2\rmi\pi}\,\sum_{k\in J}\frac{f_{q}(z)}{z-y_{k}(B)}+\oint_{|z|>\max\limits_{k\in J}y_{k}(B)}\frac{\rmd z}{2\rmi\pi}\,\sum_{k\in J}\frac{f_{q}(z)}{z-y_{k}(B)}$. The function $f_{q}$ may not have poles of order $q^{0}$ for small $q$, otherwise $A_{\p}(z)f_{q}(z)$ would not make sense as a formal Laurent series. The right side is then equal to a single contour integral around the $y_{k}(B)$, $k\in J$, which reduces to $\sum_{k\in J}f_{q}(y_{k}(B))$ by computing the residues at the $y_{k}(B)$.

In the following, singular versions of (\ref{sum[int] regular}) where $f_{q}$ has a simple pole $y_{*}=\mathcal{O}(q^{0})$ are also needed. They can be derived easily by treating separately the pure pole $b_{q}(y-y_{*})^{-1}$ with $b_{q}$ the residue of $f_{q}$ at $y_{*}$, and using (\ref{sum[int] regular}) for the regular part. For compatibility with the definitions in section~\ref{section TASEP} and the notations of \cite{P2020.2}, we subtract instead $c_{q}y(y-y_{*})^{-1}$ from $f_{q}(y)$ with $c_{q}=b_{q}/y_{*}$. For a function $f_{q}$ such that $f_{q}(y)-c_{q}y(y-y_{*})^{-1}$ is a Laurent polynomial in $y$ at each order in $q$, one has
\begin{equation}
\label{sum[int] singular}
\fl\hspace{15mm}
\sum_{k\in J}f_{q}(y_{k}(B))=c_{q}\sum_{k\in J}\frac{y_{k}(B)}{y_{k}(B)-y_{*}}+\oint\frac{\rmd z}{2\rmi\pi z}\,A_{[B,J]}(z)\Big(f_{q}(z)-\frac{c_{q}\,z}{z-y_{*}}\Big)\;.
\end{equation}
In the case $y_{*}=-\frac{N}{L-N}$, which appears frequently in the following, the term $\sum_{k\in J}\frac{y_{k}(B)}{y_{k}(B)-y_{*}}$ is in particular equal to $(L-N)\mu_{[B,J]}$ with $\mu_{\p}$ defined in (\ref{mu}). The case $y_{*}=1$ leads instead to $\eta_{\p}$ defined in (\ref{eta}).
\end{subsection}

\begin{subsection}{Integro-differential equation for the function \texorpdfstring{$Y_{\p,q}$}{Y\_pq}}
We consider in the following $g_{\p,q}$ and $Y_{\p,q}$ as the formal series in $q$ (\ref{gq[hm]}), (\ref{Y[Wm]}) with coefficients $h_{\p,m}$, $W_{\p,m}$ uniquely determined in sections \ref{section order q^1} and \ref{section order q^m} by the assumption that the $W_{\p,m}$ are Laurent polynomials. We have shown in particular that $Y_{\p,q}(y)$ depends on $\p$ only through the $\alpha_{\p,r}$. Furthermore, one has
\begin{equation}
\label{Y(q=0)}
Y_{\p,0}(y)=y
\end{equation}
and
\begin{equation}
\label{Y(y=1)}
Y_{\p,q}(1)=1
\end{equation}
for any $\p\in\R$.

We rewrite the Bethe equation (\ref{Bethe eq Y}) as
\begin{equation}
\fl\hspace{5mm}
\frac{g_{\p,q}\Pi_{\p}}{B}\,\Big(\frac{1}{1-y}\,\frac{1-Y_{\p,q}(y)}{1-q\,Y_{\p,q}(y)}\Big)^{L}=\prod_{k\in J}\Bigg(\frac{y_{k}(B)}{y}\,\frac{Y_{\p,q}(y)-q\,Y_{\p,q}(y_{k}(B))}{Y_{\p,q}(y_{k}(B))-q\,Y_{\p,q}(y)}\Bigg)\;,
\end{equation}
with $\Pi_{\p}$ defined in (\ref{Pi}). The product over $k\in J$ can be conveniently computed by residues using (\ref{sum[int] regular}) since $\log(\frac{z}{y}\,\frac{Y_{\p,q}(y)-q\,Y_{\p,q}(z)}{Y_{\p,q}(z)-q\,Y_{\p,q}(y)})$ is a Laurent polynomial in $z$ at each order in $q$ because of (\ref{Y(q=0)}). We obtain
\begin{equation}
\label{eq g Y}
\fl\hspace{2mm}
\frac{g_{\p,q}\Pi_{\p}}{B}\,\Big(\frac{1}{1-y}\,\frac{1-Y_{\p,q}(y)}{1-q\,Y_{\p,q}(y)}\Big)^{L}
=\exp\Big(\oint\frac{\rmd z}{2\rmi\pi z}\,A_{\p}(z)\log\Big(\frac{z}{y}\,\frac{Y_{\p,q}(y)-q\,Y_{\p,q}(z)}{Y_{\p,q}(z)-q\,Y_{\p,q}(y)}\Big)\Big)\;.
\end{equation}
Taking the logarithmic derivative with respect to $y$ in order to eliminate $g_{\p,q}$ finally gives a closed non-linear integro-differential equation for the function $Y_{\p,q}$,
\begin{eqnarray}
\label{eq Y}
&&\fl\hspace{2mm} \frac{N+(L-N)y}{y(1-y)}=Y_{\p,q}'(y)\,\Bigg(\frac{L}{1-Y_{\p,q}(y)}-\frac{qL}{1-qY_{\p,q}(y)}\\
&&\hspace{22mm} +\oint\frac{\rmd z}{2\rmi\pi z}\,A_{\p}(z)\Big(\frac{1}{Y_{\p,q}(y)-qY_{\p,q}(z)}+\frac{q}{Y_{\p,q}(z)-qY_{\p,q}(y)}\Big)\Bigg)\;.\nonumber
\end{eqnarray}
This equation can easily be solved perturbatively in $q$ under the assumption that $Y_{\p,q}(y)$ is a Laurent polynomial in $y$ at each order in $q$ that verifies (\ref{Y(q=0)}) and (\ref{Y(y=1)}), and we recover the results from sections~\ref{section order q^1} and \ref{section order q^m}. The expansion of $Y_{\p,q}$ up to order $q^{5}$ for systems up to size $L=7$ was checked against a numerical solution of ASEP Baxter's equation \cite{PM2008.1} for small values of $q$ giving all ${L}\choose{N}$ solutions of the Bethe equations. Perfect agreement was found.

It is convenient for the following to introduce the functions
\begin{eqnarray}
\label{U}
&& U_{\p,q}(y)=\frac{y(1-y)Y_{\p,q}'(y)}{N+(L-N)y}\\
\label{V}
&& V_{\p,q}(y)=\frac{1}{1-Y_{\p,q}(y)}-\frac{q}{1-qY_{\p,q}(y)}\;,
\end{eqnarray}
and
\begin{equation}
\label{X}
X_{\p,q}(y,z)=\frac{1}{Y_{\p,q}(y)-qY_{\p,q}(z)}+\frac{q}{Y_{\p,q}(z)-qY_{\p,q}(y)}\;.
\end{equation}
Using (\ref{sum[int] regular}), the equation (\ref{eq Y}) for $Y_{\p,q}$ can then be rewritten in terms of the functions above as
\begin{equation}
\label{eq Y UVX}
\hspace{5mm} U_{\p,q}(y)\,\Big(LV_{\p,q}(y)+\sum_{k\in J}X_{\p,q}(y,y_{k}(B))\Big)=1\;.
\end{equation}

The fugacity $g_{\p,q}$ is given in terms of $Y_{\p,q}$ by (\ref{eq g Y}), which holds for arbitrary $y$. Taking the limit $y\to1$, one has in particular
\begin{equation}
\label{g[Y]}
g_{\p,q}=\frac{B}{\Pi_{\p}}\,\frac{(1-q)^{L}}{Y_{\p,q}'(1)^{L}}\,\exp\Big(\oint\frac{\rmd z}{2\rmi\pi z}\,A_{\p}(z)\log\Big(z\,\frac{1-qY_{\p,q}(z)}{Y_{\p,q}(z)-q}\Big)\Big)\;.
\end{equation}

With the formalism developed in this section, the Bethe roots $Y_{j}=Y_{\p,q}(y_{j}(B))$ of ASEP, $\p=[B,J]$, are well defined as long as $B$ is not too close to $0$, i.e. for $|B|>\epsilon$ with $\epsilon>0$ independent of $q$. In particular, when $B\to\infty$, the solution of (\ref{eq Y UVX}) turns out to be $Y_{[\infty,J],q}(y)=\frac{y+q}{1+qy}$, and all the $Y_{j}$ converge to $1$ independently of $q$ and of the sheet index $J$. Similarly, the equation (\ref{eq Y UVX}) is perfectly regular when $B\to B_{*}$, and leads to $Y_{j}=Y_{k}$ whenever $y_{j}(B_{*})=y_{k}(B_{*})$. For small $|B|$, however, some coefficients $\alpha_{\p,m}$ appearing in the small $q$ expansion of $Y_{\p,q}$ are divergent, which implies that the limits $q\to0$ and $B\to0$ do not commute, and our approach breaks down. This is however not a problem for writing the probability of the height, at least for a finite system, as contour integrals avoiding the points $[0,J]$ can be used in section~\ref{section proba} below.
\end{subsection}
\end{section}

\begin{section}{Probability of the height}
\label{section proba}
In this section we compute the probability of the height for ASEP in terms of the function $Y_{\p,q}$ introduced in the previous section, with a focus on the evolution with stationary initial condition, for which the scalar product between eigenstates of $M_{i}(q,g_{\p,q})$ and the initial state has a particularly simple expression.

\begin{subsection}{Generating function of the height}
We consider ASEP with $N$ particles on $L$ sites and periodic boundary conditions. We recall in the following the known mapping from ASEP to an interface growth model, and the deformed Markov matrix formalism in terms of which the generating function of the interface height can be expressed.

For any configuration $\mathcal{C}$ of the particles and $i\in\[1,L\]$, we define
\begin{equation}
\label{Hi(C)}
\mathcal{H}_{i}(\mathcal{C})=\sum_{\ell=1}^{i}\Big(\frac{N}{L}-n_{\ell}(\mathcal{C})\Big)\;,
\end{equation}
where $n_{i}(\mathcal{C})$ is the occupation number at site $i$ for the configuration $\mathcal{C}$, with $n_{i}(\mathcal{C})=1$ corresponding to a site occupied by a particle and $n_{i}(\mathcal{C})=0$ to an empty site. This definition is then extended to any $i\in\Z$ by periodicity, $\mathcal{H}_{i+L}(\mathcal{C})=\mathcal{H}_{i}(\mathcal{C})$. One has $\mathcal{H}_{i+1}(\mathcal{C})=\mathcal{H}_{i}(\mathcal{C})\pm1$ for any site $i\in\mathbb{Z}$.

\begin{figure}
	\begin{center}
		\begin{picture}(120,70)
			\newcommand{\Up}[1]{\put #1{\line(1,1){10}}}
			\newcommand{\Down}[1]{\put #1{\line(1,-1){10}}}
			\newcommand{\UpThick}[1]{\put #1{\thicklines\line(1,1){10}}}
			\newcommand{\DownThick}[1]{\put #1{\thicklines\line(1,-1){10}}}
			\newcommand{\UpDotted}[1]{\put #1{\color[rgb]{0.7,0.7,0.7}\line(1,1){10}\color{black}}}
			\newcommand{\DownDotted}[1]{\put #1{\color[rgb]{0.7,0.7,0.7}\line(1,-1){10}\color{black}}}
			\put(5,3){\circle*{2}}\put(25,3){\circle*{2}}\put(35,3){\circle*{2}}\put(75,3){\circle*{2}}\put(85,3){\circle*{2}}\put(95,3){\circle*{2}}\put(115,3){\circle*{2}}
			\qbezier(5,5)(10,10)(15,5)\put(15,5){\vector(1,-1){0.2}}\put(8,9){$1$}\qbezier(65,5)(70,10)(75,5)\put(65,5){\vector(-1,-1){0.2}}\put(68,10){$q$}
			\put(0,0){\line(1,0){120}}\multiput(0,0)(10,0){13}{\line(0,1){5}}\UpDotted{(0,20)}\DownDotted{(10,30)}\UpDotted{(20,20)}\DownThick{(30,30)}\UpThick{(40,20)}\DownDotted{(50,30)}\UpDotted{(60,20)}\DownDotted{(70,30)}\UpDotted{(80,20)}\DownThick{(90,30)}\UpThick{(100,20)}\DownDotted{(110,30)}\DownThick{(0,40)}\UpThick{(10,30)}\DownThick{(20,40)}\UpThick{(50,30)}\DownThick{(60,40)}\UpThick{(70,30)}\DownThick{(80,40)}\UpThick{(110,30)}\put(10,47){\vector(0,-1){14}}\put(11,39){$1$}\Down{(0,60)}\Up{(10,50)}\Up{(0,60)}\Down{(10,70)}\put(70,55){\vector(0,1){14}}\put(71,60){$q$}\Down{(60,42)}\Up{(70,32)}\Up{(60,42)}\Down{(70,52)}
		\end{picture}
	\end{center}
	\caption{Mapping between ASEP and an interface growth model.}
	\label{fig mapping height}
\end{figure}
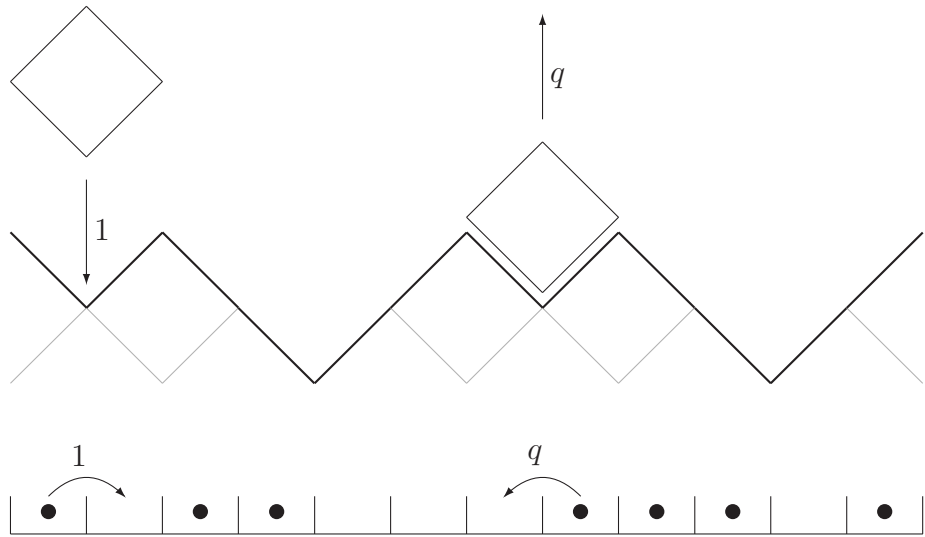

We call $\mathcal{C}_{t}$ the configuration of the system at time $t$, and we associate to the ASEP dynamics a height function $H_{i}(t)$, $i\in\Z$. The initial height at time $t=0$ is defined by $H_{i}(0)=\mathcal{H}_{i}(\mathcal{C}_{0})$. The height $H_{i}$ is then updated every time a particle moves between sites $i$ and $i+1$ (counted modulo $L$), with $H_{i}\to H_{i}+1$ when a particle hops from site $i$ to $i+1$ (which happens at rate $1$) and $H_{i}\to H_{i}-1$ when a particle hops from site $i+1$ to $i$ (which happens at rate $q$), see figure~\ref{fig mapping height}. With this dynamics, $H_{i}(t)=H_{0}(t)+\mathcal{H}_{i}(\mathcal{C}_{t})$ for any $i\in\[1,L\]$, periodicity $H_{i+L}(t)=H_{i}(t)$ is preserved, and $H_{i+1}(t)=H_{i}(t)\pm1$ holds at any time $t\geq0$. The height function is closely related to the time-integrated current of particles: indeed, the difference $H_{i}(t)-H_{i}(0)$ is equal to the total number of times particles have hopped from site $i$ to $i+1$ minus the total number of times particles have hopped from site $i+1$ to $i$.

Considering for the moment that the system is prepared initially in the configuration $\mathcal{C}_{0}$, the Markov matrix $M(q)$ of ASEP gives the probability $P_{t}(\mathcal{C})$ for the system to be in configuration $\mathcal{C}$ at time $t$ as
\begin{equation}
P_{t}(\mathcal{C})=\langle\mathcal{C}|\rme^{tM(q)}|\mathcal{C}_{0}\rangle\;.
\end{equation}
The generating function of the height (or, equivalently, of the current of particles) can be computed using a deformation $M_{i}(q,\rme^{\gamma})$ of the Markov matrix, built by multiplying the non-diagonal entries of $M(q)$ corresponding to particles hopping from site $i$ to $i+1$ by $\rme^{\gamma}$ and those corresponding to particles hopping from site $i+1$ to $i$ by $\rme^{-\gamma}$, see e.g. \cite{DL1998.1,PM2008.1,LM2011.1}. One has
\begin{equation}
\label{GF[Mi]}
\langle\rme^{\gamma(H_{i}(t)-H_{i}(0))}\rangle_{\mathcal{C}_{0}}=\sum_{\mathcal{C}\in\Omega}\langle\mathcal{C}|\rme^{tM_{i}(q,\rme^{\gamma})}|\mathcal{C}_{0}\rangle\;,
\end{equation}
where the expectation value on the left is over all histories starting in the configuration $\mathcal{C}_{0}$, and the summation on the right is over the set $\Omega$ of all ${L}\choose{N}$ possible configurations of the particles in the system verifying the exclusion constraint.
\end{subsection}

\begin{subsection}{Expansion over eigenstates}
In preparation for the eigenstate expansion, we rewrite the generating function of the height in terms of a translation invariant version of the deformed Markov matrix. The derivation is essentially the same as for TASEP, see \cite{P2020.2}.

The local fugacity at site $i$ of the deformed Markov matrix $M_{i}(q,\rme^{\gamma})$ can be spread uniformly on the whole system by a similarity transformation. Defining the globally deformed, translation invariant Markov matrix $M(q,\rme^{\gamma/L})$, obtained by multiplying non-diagonal entries of the non-deformed Markov matrix $M(q)$ by $\rme^{\gamma/L}$ or $\rme^{-\gamma/L}$ depending respectively on whether the entries correspond to particles moving forward or backward, one has
\begin{equation}
\label{Mi[M]}
M_{i}(q,\rme^{\gamma})=\rme^{-\gamma S_{i}}M(q,\rme^{\gamma/L})\,\rme^{\gamma S_{i}}\;,
\end{equation}
with $S_{i}$ defined by
\begin{equation}
\label{Si}
S_{i}|\mathcal{C}\rangle=\Big(\frac{1}{L}\sum_{j=1}^{N}x_{j}^{(i)}\Big)|\mathcal{C}\rangle\;,
\end{equation}
where the $x_{j}^{(i)}\in\[1,L\]$ are the positions of the particles counted from site $i$ (i.e. a particle at site $i+\ell$ modulo $L$ corresponds to $x_{j}^{(i)}=\ell$) in the configuration $\mathcal{C}$. We note that the operator $S_{i}$ is related to the function $\mathcal{H}_{i}$ defined in (\ref{Hi(C)}) by $(S_{0}-S_{i})|\mathcal{C}\rangle=\mathcal{H}_{i}(\mathcal{C})|\mathcal{C}\rangle$.

The translation operator $T$, acting in configuration basis as $T|x_{1},\ldots,x_{N}\rangle=|x_{1}-1,\ldots,x_{N}-1\rangle$ with configurations written in terms of the positions $x_{j}$ of the particles, and such that $T^{L}=1$, commutes with $M(q,\rme^{\gamma/L})$. Furthermore, one has $S_{i}=T^{-i}S_{0}T^{i}$ and $\sum_{\mathcal{C}\in\Omega}\langle\mathcal{C}|T=\sum_{\mathcal{C}\in\Omega}\langle\mathcal{C}|$.

After some rewriting, and cancellation of the factor $\rme^{-\gamma H_{i}(0)}=\rme^{-\gamma\mathcal{H}_{i}(\mathcal{C}_{0})}$ on both sides, one has from (\ref{GF[Mi]}) $\langle\rme^{\gamma H_{i}(t)}\rangle_{\mathcal{C}_{0}}
=\sum_{\mathcal{C}\in\Omega}\langle\mathcal{C}|\rme^{-\gamma S_{0}}T^{i}\rme^{tM(q,\rme^{\gamma/L})}\rme^{\gamma S_{0}}|\mathcal{C}_{0}\rangle$. At this point, we can finally consider a more general initial state at time $t=0$, with a random initial configuration distributed according to the probabilities $P_{0}(\mathcal{C})\in[0,1]$, normalized as $\sum_{\mathcal{C}\in\Omega}P_{0}(\mathcal{C})=1$. Introducing the vector $|P_{0}\rangle=\sum_{\mathcal{C}\in\Omega}P_{0}(\mathcal{C})|\mathcal{C}\rangle$, we obtain
\begin{equation}
\label{GF[M] P0}
\langle\rme^{\gamma H_{i}(t)}\rangle
=\sum_{\mathcal{C}\in\Omega}\langle\mathcal{C}|\rme^{-\gamma S_{0}}T^{i}\rme^{tM(q,\rme^{\gamma/L})}\rme^{\gamma S_{0}}|P_{0}\rangle\;,
\end{equation}
where the averaging is now over all histories with initial probabilities $P_{0}(\mathcal{C})$.

We index the eigenstates of the various Markov matrices by the sets $J$ labelling the solutions of the Bethe equations introduced in section~\ref{section Y}, and call $\langle\psi_{J}(q,\gamma)|$ and $|\psi_{J}(q,\gamma)\rangle$ (respectively $\langle\psi_{J}^{0}(q,\gamma)|$ and $|\psi_{J}^{0}(q,\gamma)\rangle$) the left and right eigenvectors of $M(q,\rme^{\gamma/L})$ (resp. $M_{0}(q,\rme^{\gamma})$). From (\ref{Mi[M]}), one has $\langle\psi_{J}^{0}(q,\gamma)|=\langle\psi_{J}(q,\gamma)|\rme^{\gamma S_{0}}$ and $|\psi_{J}^{0}(q,\gamma)\rangle=\rme^{-\gamma S_{0}}|\psi_{J}(q,\gamma)\rangle$. The matrices $M(q,\rme^{\gamma/L})$ and $M_{0}(q,\rme^{\gamma})$ have the same eigenvalues, that we call $E_{J}(q,\gamma)$. Additionally, since the matrix $M(q,\rme^{\gamma/L})$ commutes with the translation operator $T$, both $\langle\psi_{J}(q,\gamma)|$ and $|\psi_{J}(q,\gamma)\rangle$ are eigenvectors of $T$ with eigenvalue $\rme^{\rmi P_{J}/L}$, $P_{J}\in2\pi\Z$. We finally obtain the eigenstate expansion of (\ref{GF[M] P0}) as
\begin{equation}
\label{GF[psi0]}
\langle\rme^{\gamma H_{i}(t)}\rangle
=\sum_{J}\rme^{tE_{J}(q,\gamma)}\,\rme^{\rmi P_{J}\,i/L}\,
\frac{\sum_{\mathcal{C}\in\Omega}\langle\mathcal{C}|\psi_{J}^{0}(q,\gamma)\rangle\langle\psi_{J}^{0}(q,\gamma)|P_{0}\rangle}{\langle\psi_{J}^{0}(q,\gamma)|\psi_{J}^{0}(q,\gamma)\rangle}\;.
\end{equation}

For an evolution with stationary initial condition, corresponding for ASEP to a uniform distribution for the initial configurations, this leads to
\begin{equation}
\label{GF[psi0] stat}
\fl\hspace{10mm}
\langle\rme^{\gamma H_{i}(t)}\rangle_{\text{stat}}
=\frac{1}{|\Omega|}\sum_{J}\rme^{tE_{J}(q,\gamma)}\,\rme^{\rmi P_{J}\,i/L}\,
\frac{\sum_{\mathcal{C}\in\Omega}\langle\mathcal{C}|\psi_{J}^{0}(q,\gamma)\rangle\;\sum_{\mathcal{C}_{0}\in\Omega}\langle\psi_{J}^{0}(q,\gamma)|\mathcal{C}_{0}\rangle}{\langle\psi_{J}^{0}(q,\gamma)|\psi_{J}^{0}(q,\gamma)\rangle}\;.
\end{equation}
\end{subsection}

\begin{subsection}{Bethe ansatz formulas}
The left and right eigenvectors $\langle\psi_{J}^{0}(q,\gamma)|$ and $|\psi_{J}^{0}(q,\gamma)\rangle$ of the deformed Markov matrix $M_{0}(q,\rme^{\gamma})$ can be expressed in terms of the corresponding solution $\{Y_{j},j\in J\}$ of the Bethe equations (\ref{Bethe eq ASEP}) with $g=\rme^{\gamma}$. The Bethe ansatz formulas given in this section are quoted from \cite{P2016.2}, with $\gamma$ replaced by $\gamma/L$ and a slightly altered normalization for the eigenvectors in order to get simpler formulas.

The eigenvalue of the deformed Markov matrices $M(q,\rme^{\gamma/L})$ and $M_{i}(q,\rme^{\gamma})$ corresponding to the Bethe roots $Y_{j}$, $j\in J$ is equal to
\begin{equation}
E_{J}(q,\gamma)=(1-q)\sum_{j\in J}\Big(\frac{1}{1-Y_{j}}-\frac{1}{1-qY_{j}}\Big)\;,
\end{equation}
and the corresponding eigenvalue for the translation operator $T$ is
\begin{equation}
\rme^{\rmi P_{J}/L}=\rme^{N\gamma/L}\prod_{j\in J}\frac{1-Y_{j}}{1-qY_{j}}\;.
\end{equation}

The entries of the left and right eigenvectors of $M_{0}(q,\rme^{\gamma})$ for a configuration $\mathcal{C}$ corresponding to particles at positions $x_{j}$, $1\leq x_{1}<\ldots<x_{N}\leq L$ are given by symmetric functions of the Bethe roots,
\begin{equation}
\label{<C|psi0>}
\fl\hspace{10mm}
\langle\mathcal{C}|\psi_{J}^{0}(q,\gamma)\rangle
=\sum_{\sigma\in S_{N}}\prod_{j=1}^{N}
\Bigg(
	\frac{1}{1-\tilde{Y}_{\sigma(j)}}\,
	\Big(\frac{1-\tilde{Y}_{\sigma(j)}}{1-q\tilde{Y}_{\sigma(j)}}\Big)^{x_{j}}
	\prod_{k=j+1}^{N}\frac{\tilde{Y}_{\sigma(j)}-q\tilde{Y}_{\sigma(k)}}{\tilde{Y}_{\sigma(j)}-\tilde{Y}_{\sigma(k)}}
\Bigg)
\end{equation}
and
\begin{equation}
\label{<psi0|C>}
\fl\hspace{10mm}
\langle\psi_{J}^{0}(q,\gamma)|\mathcal{C}\rangle
=\sum_{\sigma\in S_{N}}\prod_{j=1}^{N}\Bigg(
\frac{1}{1-\tilde{Y}_{\sigma(j)}}\,
\Big(\frac{1-\tilde{Y}_{\sigma(j)}}{1-q\tilde{Y}_{\sigma(j)}}\Big)^{L+1-x_{j}}
\prod_{k=j+1}^{N}\frac{\tilde{Y}_{\sigma(k)}-q\tilde{Y}_{\sigma(j)}}{\tilde{Y}_{\sigma(k)}-\tilde{Y}_{\sigma(j)}}
\Bigg)\;,
\end{equation}
where we used here the notation $\tilde{Y}_{j}$, $j=1,\ldots,N$ for the Bethe roots instead of $Y_{j}$, $j\in J$ for convenience. The summation is over the group $S_{N}$ of permutations of $N$ elements. These eigenstates are normalized as
\begin{eqnarray}
\label{<psi0|psi0>}
&&\fl\hspace{2mm}\langle\psi_{J}^{0}(q,\gamma)|\psi_{J}^{0}(q,\gamma)\rangle=\frac{(-1)^{N}}{(1-q)^{N}\rme^{N\gamma}}\,\Bigg(\prod_{j,k\in J\atop j<k}\frac{(Y_{j}-qY_{k})(qY_{j}-Y_{k})}{(Y_{j}-Y_{k})^{2}}\Bigg)\nonumber\\
&&\hspace{20mm} \times\det\Bigg(\partial_{Y_{i}}\log\Big(\Big(\frac{1-Y_{j}}{1-qY_{j}}\Big)^{L}\prod_{k\in J}\frac{qY_{j}-Y_{k}}{Y_{j}-qY_{k}}\Big)\Bigg)_{i,j\in J}\;.
\end{eqnarray}
The determinant of a derivative with respect to the Bethe roots of the Bethe equations in logarithmic form found in (\ref{<psi0|psi0>}), known as the Gaudin determinant \cite{GMCW1981.1,S1989.1}, appears for the normalization of eigenstates in various quantum integrable models.

From (\ref{GF[psi0]}), the scalar product between $\sum_{\mathcal{C}\in\Omega}\langle\mathcal{C}|$ and right eigenvectors is needed. For stationary initial condition (\ref{GF[psi0] stat}), the scalar product between left eigenvectors and $\sum_{\mathcal{C}\in\Omega}|\mathcal{C}\rangle$ is also needed. With the eigenvectors normalized as in (\ref{<C|psi0>}), (\ref{<psi0|C>}), one has from \cite{P2016.2}
\begin{equation}
\label{<sum|psi0>}
\sum_{\mathcal{C}\in\Omega}\langle\mathcal{C}|\psi_{J}^{0}(q,\gamma)\rangle
=\sum_{\mathcal{C}\in\Omega}\langle\psi_{J}^{0}(q,\gamma)|\mathcal{C}\rangle
=\Big(\prod_{j\in J}Y_{j}^{-1}\Big)\prod_{j=0}^{N-1}\frac{1-\rme^{-\gamma}\,q^{j}}{1-q}\;.
\end{equation}

The generating function of the height (\ref{GF[psi0] stat}) can finally be expressed in terms of the Bethe roots as
\begin{eqnarray}
\label{GF[Yj]}
&&\fl\hspace{5mm}
\langle\rme^{\gamma H_{i}(t)}\rangle=(-1)^{N}\rme^{N\gamma}\Bigg(\prod_{j=0}^{N-1}(1-\rme^{-\gamma}\,q^{j})\Bigg)\,
\sum_{J}\langle\psi_{J}^{0}(q,\gamma)|P_{0}\rangle\,\rme^{(1-q)t\sum\limits_{j\in J}\big(\frac{1}{1-Y_{j}}-\frac{1}{1-qY_{j}}\big)}\\
&& \times\Bigg(\rme^{N\gamma/L}\prod_{j\in J}\frac{1-Y_{j}}{1-qY_{j}}\Bigg)^{i}\,
\frac{\Big(\prod_{j\in J}Y_{j}^{-1}\Big)\Big(\prod_{j,k\in J\atop j<k}\frac{(Y_{j}-Y_{k})^{2}}{(Y_{j}-qY_{k})(qY_{j}-Y_{k})}\Big)}{\det\Bigg(\partial_{Y_{i}}\log\Big(\Big(\frac{1-Y_{j}}{1-qY_{j}}\Big)^{L}\prod_{k\in J}\frac{qY_{j}-Y_{k}}{Y_{j}-qY_{k}}\Big)\Bigg)_{i,j\in J}}\;,\nonumber
\end{eqnarray}
where $\{Y_{j},j\in J\}$ is the solution of the Bethe equations (\ref{Bethe eq ASEP}) with $g=\rme^{\gamma}$ corresponding to the label $J$. For stationary initial condition, the scalar product with the initial state reduces from (\ref{<sum|psi0>}) to $\langle\psi_{J}^{0}(q,\gamma)|P_{0}\rangle=\frac{\prod_{j\in J}Y_{j}^{-1}}{{{L}\choose{N}}}\prod_{j=0}^{N-1}\frac{1-\rme^{-\gamma}\,q^{j}}{1-q}$.
\end{subsection}

\begin{subsection}{Function \texorpdfstring{$Y_{\p,q}$}{Y\_pq} and contour integrals}
In the next step, we finally use the perturbative solution of the Bethe equations developed in section~\ref{section Y}. We rewrite all symmetric functions of the Bethe roots appearing in the generating function (\ref{GF[Yj]}) in terms of contour integrals involving the solution $Y_{\p,q}(y)$, $\p\in\R$ of the integro-differential equation (\ref{eq Y}).

\begin{subsubsection}{Products and sums}\hfill\\
For simple products of the form $\prod_{j\in J}f_{q}(y_{j}(B))$ with $f_{q}(y)=\sum_{m=0}^{\infty}c_{m}(y)q^{m}$ a formal series in $q$ with $c_{0}(y)=1$ and whose other coefficients $c_{m}(y)$ are Laurent polynomials in $y$, the regular summation formula (\ref{sum[int] regular}) can be used directly, and one has
\begin{equation}
\prod_{j\in J}f_{q}(y_{j}(B))=\exp\Big(\oint\frac{\rmd z}{2\rmi\pi z}\,A_{[B,J]}(z)\log(f_{q}(z))\Big)\;.
\end{equation}
The constraints above for the coefficients $c_{m}(y)$ ensure that the product $A_{[B,J]}(z)\log(f_{q}(z))$ is well defined.

For the product $\prod_{j\in J}Y_{j}=\Pi_{[B,J]}\,\prod_{j\in J}\frac{Y_{j}}{y_{j}(B)}$ where $\Pi_{\p}$ is defined in (\ref{Pi}), writing $Y_{j}=Y_{[B,J],q}(y_{j}(B))$, one has in particular
\begin{equation}
\label{prodY[int]}
\prod_{j\in J}Y_{j}=\Pi_{\p}\,\exp\Big(\oint\frac{\rmd z}{2\rmi\pi z}\,A_{\p}(z)\,\log\Big(\frac{Y_{\p,q}(z)}{z}\Big)\Big)\;,
\end{equation}
with $\p=[B,J]\in\R$. The product $\prod_{j\in J}\frac{1-Y_{j}}{1-qY_{j}}=\bar{\Pi}_{[B,J]}\,\prod_{j\in J}\frac{1-Y_{j}}{(1-y_{j}(B))(1-qY_{j})}$ where $\bar{\Pi}_{\p}$ is defined in (\ref{Pib}) can be treated in the same way, and one finds
\begin{equation}
\label{prodMomentum[int]}
\prod_{j\in J}\frac{1-Y_{j}}{1-qY_{j}}=\bar{\Pi}_{\p}\,\exp\Big(\oint\frac{\rmd z}{2\rmi\pi z}\,A_{\p}(z)\,\log\Big(\frac{1}{1-z}\,\frac{1-Y_{\p,q}(z)}{1-qY_{\p,q}(z)}\Big)\Big)\;,
\end{equation}
with $\p=[B,J]\in\R$.

The double product in (\ref{GF[Yj]}) can be treated in a similar way, and we obtain
\begin{eqnarray}
\label{prodprod[int]}
&&\fl\hspace{2mm} \prod_{j,k\in J\atop j<k}\frac{(Y_{j}-Y_{k})^{2}}{(Y_{j}-qY_{k})(qY_{j}-Y_{k})}
=\frac{(-1)^{\frac{N(N-1)}{2}}\,(1-q)^{N}\,V^{2}_{\p}}{\Pi_{\p}^{N-1}}\nonumber\\
&&\hspace{20mm}
\times\exp\Big(\oint\frac{\rmd z}{2\rmi\pi z}\,A_{\p}(z)\log\Big(\frac{Y_{\p,q}(z)}{z\,Y_{\p,q}'(z)}\Big)\Big)\\
&&\hspace{15mm}
\times\exp\Big(\frac{1}{2}\oint\frac{\rmd w}{2\rmi\pi w}\oint\frac{\rmd z}{2\rmi\pi z}\,A_{\p}(w)A_{\p}(z)\nonumber\\
&&\hspace{27mm}
\log\Big(\frac{w z}{(w-z)^{2}}\,\frac{(Y_{\p,q}(w)-Y_{\p,q}(z))^{2}}{(Y_{\p,q}(w)-qY_{\p,q}(z))(Y_{\p,q}(z)-qY_{\p,q}(w))}\Big)\Big)\;,\nonumber
\end{eqnarray}
with $\p=[B,J]\in\R$, $V^{2}_{\p}$ the squared Vandermonde determinant defined in (\ref{V2}) and $\Pi_{\p}$ the simple product defined in (\ref{Pi}). The singularity at $z=w$ cancels in the double integral: the logarithm is a Laurent polynomial in both variables $w$ and $z$ at each order in $q$, and the ordering of the contours of integration does not matter.

For the sum $\sum_{j\in J}(\frac{1}{1-Y_{j}}-\frac{1}{1-qY_{j}})$ coming from the eigenvalue, the singular summation formula (\ref{sum[int] singular}) with $y_{*}=1$ has to be used due to the pole at $Y_{j}=1$. The residue at $y=1$ of $\frac{1}{1-Y_{\p,q}(y)}-\frac{1}{1-qY_{\p,q}(y)}$ is equal to $-1/Y_{\p,q}'(1)$. Writing again $\p=[B,J]\in\R$, we obtain
\begin{eqnarray}
\label{E[int]}
&&\fl\hspace{2mm}
\sum_{j\in J}\Big(\frac{1}{1-Y_{j}}-\frac{1}{1-qY_{j}}\Big)=\frac{\eta_{\p}}{Y_{\p,q}'(1)}+\oint\frac{\rmd z}{2\rmi\pi z}\,A_{\p}(z)\,\Big(\frac{1}{1-Y_{\p,q}(z)}-\frac{1}{1-qY_{\p,q}(z)}\\
&&\hspace{91mm}
-\frac{1}{Y_{\p,q}'(1)}\,\frac{z}{1-z}\Big)\;,\nonumber
\end{eqnarray}
where $\eta_{\p}$ is defined in (\ref{eta}). The integrand is again equal to $A_{\p}(z)$ multiplied by a formal series in $q$ whose coefficients are Laurent polynomials in $z$, so that the contour integral does make sense.
\end{subsubsection}

\begin{subsubsection}{Initial condition}\hfill\\
From (\ref{GF[Yj]}), the scalar product between left eigenvectors and the vector $|P_{0}\rangle$ representing the probabilities of initial configurations of the particles is needed, and we introduce for $\p=[B,J]$ the notation
\begin{equation}
\label{Theta}
\Theta_{\p,q}^{P_{0}}=\langle\psi_{J}^{0}(q,\log g_{\p,q})|P_{0}\rangle=\sum_{\mathcal{C}\in\Omega}P_{0}(\mathcal{C})\,\langle\psi_{J}^{0}(q,\log g_{\p,q})|\mathcal{C}\rangle\;,
\end{equation}
where the scalar product is defined in terms of the Bethe roots $Y_{j}=Y_{\p,q}(y_{j}(B))$ by (\ref{<psi0|C>}). For stationary initial condition, one has from (\ref{<sum|psi0>}) the particularly simple expression
\begin{equation}
\Theta_{\p,q}^{\text{stat}}=\frac{\prod_{j\in J}Y_{j}^{-1}}{{{L}\choose{N}}}\,\prod_{j=0}^{N-1}\frac{1-q^{j}/g_{\p,q}}{1-q}\;,
\end{equation}
where $\prod_{j\in J}Y_{j}^{-1}$ and $g_{\p,q}$ have explicit contour integral expressions (\ref{g[Y]}), (\ref{prodY[int]}).
\end{subsubsection}

\begin{subsubsection}{Gaudin determinant}\hfill\\
The last piece of the expression (\ref{GF[Yj]}) for the generating function depending on the Bethe roots $Y_{j}$ is the Gaudin determinant
\begin{equation}
D_{\p,q}=\det\Bigg(\partial_{Y_{i}}\log\Big(\Big(\frac{1-Y_{j}}{1-qY_{j}}\Big)^{L}\prod_{k\in J}\frac{qY_{j}-Y_{k}}{Y_{j}-qY_{k}}\Big)\Bigg)_{i,j\in J}\;,
\end{equation}
where $\p=[B,J]$ and all $Y_{j}$'s are set equal to $Y_{\p,q}(y_{j}(B))$ after taking the derivative with respect to $Y_{i}$. Computing explicitly the derivatives, one has in terms of $X_{\p,q}$ defined in (\ref{X}) and $V_{\p,q}$ defined in (\ref{V})
\begin{equation}
\fl
D_{\p,q}=\det\Big(X_{\p,q}(y_{i}(B),y_{j}(B))
-\delta_{i,j}\Big(LV_{\p,q}(y_{i}(B))+\sum_{k\in J}X_{\p,q}(y_{i}(B),y_{k}(B))\Big)\Big)_{i,j\in J}\;.\nonumber
\end{equation}
From (\ref{eq Y UVX}), the term with the Kronecker delta reduces to $-\delta_{i,j}/U_{\p,q}(y_{i}(B))$, with $U_{\p,q}$ defined in (\ref{U}). Then, one has
\begin{equation}
\fl\hspace{10mm}
D_{\p,q}=\frac{(-1)^{N}}{\prod_{k\in J}U_{\p,q}(y_{k}(B))}\,
\det\Big(\delta_{i,j}-U_{\p,q}(y_{i}(B))\,X_{\p,q}(y_{i}(B),y_{j}(B))\Big)_{i,j\in J}\;.
\end{equation}
It is convenient to subtract $X_{\p,q}(y_{i}(B),-\tfrac{N}{L-N})$ in order to eventually use the regular summation formula (\ref{sum[int] regular}) for the determinant. Defining
\begin{equation}
\label{K}
K_{\p,q}(y,z)=U_{\p,q}(y)\,\Big(X_{\p,q}(y,z)-X_{\p,q}(y,-\tfrac{N}{L-N})\Big)\;,
\end{equation}
one has
\begin{eqnarray}
&&
D_{\p,q}=\frac{(-1)^{N}}{\prod_{k\in J}U_{\p,q}(y_{k}(B))}\,
\det\Big(\delta_{i,j}-K_{\p,q}(y_{i}(B),y_{j}(B))\\
&&\hspace{60mm}
-U_{\p,q}(y_{i}(B))X_{\p,q}(y_{i}(B),-\tfrac{N}{L-N})\Big)_{i,j\in J}\;.\nonumber
\end{eqnarray}
Using the general formula for rank one perturbation of determinants, $\det(A+BC^{\top})=\det(A)\,(1+C^{\top}A^{-1}B)$ with $A$ a square matrix and $B$, $C$ column vectors, we obtain
\begin{eqnarray}
&&
D_{\p,q}=\frac{(-1)^{N}\,\det(\mathrm{Id}-\mathrm{K}_{\p,q})}{\prod_{k\in J}U_{\p,q}(y_{k}(B))}\\
&&\hspace{13mm}
\times\Bigg(1-\sum_{i,j\in J}((\mathrm{Id}-\mathrm{K}_{\p,q})^{-1})_{i,j}\,U_{\p,q}(y_{j}(B))X_{\p,q}(y_{j}(B),-\tfrac{N}{L-N})\Bigg)\;,\nonumber
\end{eqnarray}
where $\mathrm{Id}$ is the $N\times N$ identity matrix and $\mathrm{K}_{\p,q}=(K_{\p,q}(y_{i}(B),y_{j}(B)))_{i,j\in J}$. Writing
\begin{eqnarray}
&&\fl
U_{\p,q}(y_{j}(B))X_{\p,q}(y_{j}(B),-\tfrac{N}{L-N})=\frac{1}{N}\sum_{k\in J}U_{\p,q}(y_{j}(B))X_{\p,q}(y_{j}(B),-\tfrac{N}{L-N})\\
&&\hspace{5mm}
=\frac{1}{N}\sum_{k\in J}\Big(U_{\p,q}(y_{j}(B))X_{\p,q}(y_{j}(B),y_{k}(B))-K_{\p,q}(y_{j}(B),y_{k}(B))\Big)\nonumber\\
&&\hspace{5mm}
=-\frac{L\,U_{\p,q}(y_{j}(B))V_{\p,q}(y_{j}(B))}{N}+\frac{1}{N}\sum_{k\in J}\Big(\delta_{j,k}-K_{\p,q}(y_{j}(B),y_{k}(B))\Big)\;,\nonumber
\end{eqnarray}
where the last equality follows from (\ref{eq Y UVX}), we finally obtain after simplifications
\begin{equation}
\label{det[int]}
D_{\p,q}=\frac{(-1)^{N}\,\det(1-\mathbb{K}_{\p,q})}{\prod_{k\in J}U_{\p,q}(y_{k}(B))}\,
\Big(\frac{L}{N}\,\langle1|(1-\mathbb{K}_{\p,q})^{-1}\,U_{\p,q}V_{\p,q}\rangle_{\p}\Big)\;,\nonumber
\end{equation}
where the scalar product of two meromorphic functions at the point $\p\in\R$ is defined as $\langle f|g\rangle_{\p}=\sum_{k\in J}f(y_{k}(B))\,g(y_{k}(B))$. The operator $\mathbb{K}_{\p,q}$ acts on (formal series in $q$ of) meromorphic functions as
\begin{equation}
\label{K op}
(\mathbb{K}_{\p,q}f)(y)=\sum_{k\in J}\,K_{\p,q}(y,y_{k}(B))\,f(y_{k}(B))\;,
\end{equation}
and using (\ref{sum[int] regular}) reduces for (formal series in $q$ of) Laurent polynomials to
\begin{equation}
(\mathbb{K}_{\p,q}f)(y)=\oint\frac{\rmd z}{2\rmi\pi z}\,A_{\p}(z)\,K_{\p,q}(y,z)\,f(z)\;.
\end{equation}
The Fredholm determinant is in particular equal through $\log\det(1-\mathbb{K}_{\p,q})=\tr\log(1-\mathbb{K}_{\p,q})$ to
\begin{equation}
\fl\hspace{15mm}
\det(1-\mathbb{K}_{\p,q})
=\exp\Bigg(-\sum_{n=1}^{\infty}\frac{1}{n}\oint\rmd z_{1}\ldots\rmd z_{n}\prod_{i=1}^{n}\Big(\frac{A_{\p}(z_{i})}{2\rmi\pi z_{i}}K_{\p,q}(z_{i+1},z_{i})\Big)\Bigg)
\end{equation}
with the convention $z_{n+1}=z_{1}$.

The function $(1-\mathbb{K}_{\p,q})^{-1}\,U_{\p,q}V_{\p,q}=\sum_{m=0}^{\infty}(\mathbb{K}_{\p,q})^{m}U_{\p,q}V_{\p,q}$ can be expressed in terms of contour integrals using the regular summation formula (\ref{sum[int] regular}), and thus depends on $\p$ only through the coefficients $\alpha_{\p,m}$ defined in (\ref{alpham}), whose poles are located at points $\p=[0,J]\in\R$. Computing the scalar product $\langle1|(1-\mathbb{K}_{\p,q})^{-1}\,U_{\p,q}V_{\p,q}\rangle_{\p}$ requires however the singular summation formula (\ref{sum[int] singular}) with a pole at $y_{*}=-\frac{N}{L-N}$, and thus gives an additional coefficient $\mu_{\p}$, defined in (\ref{mu}), which has additional poles of the form $\p=[B_{*}\pm\rmi0^{+},J]\in\R$ for some values of $J$. We show in the next section that the factor $\langle1|(1-\mathbb{K}_{\p,q})^{-1}\,U_{\p,q}V_{\p,q}\rangle_{\p}$ from the Gaudin determinant cancels when considering the probability of the height, which leads to a simple pole structure for the integrand of (\ref{P(Hi) P0}) below. The same phenomenon was already observed for TASEP in \cite{P2020.2}.
\end{subsubsection}
\end{subsection}

\begin{subsection}{Probability of the height}
\label{section P(H)}
We are interested in the probability distribution of the height, which can be expressed in terms of the generating function considered in the previous sections. Indeed, from $H_{i}(t)-H_{i}(0)\in\Z$ and the definition (\ref{Hi(C)}) of $H_{i}(0)$, one has $H_{i}(t)\in\frac{Ni}{L}+\Z$, which implies
\begin{equation}
\langle\rme^{\gamma H_{i}(t)}\rangle=\sum_{U\in\Z}\P(H_{i}(t)=\tfrac{Ni}{L}+U)\,\rme^{\gamma(\frac{Ni}{L}+U)}\;.
\end{equation}
Writing $g=\rme^{\gamma}$, the probability can then be extracted as
\begin{equation}
\label{P[GF]}
\P(H_{i}(t)=\tfrac{Ni}{L}+U)=\oint\frac{\rmd g}{2\rmi\pi g^{\frac{Ni}{L}+U+1}}\,\langle g^{H_{i}(t)}\rangle\;,
\end{equation}
with $U\in\Z$ and an integration over a counter-clockwise contour encircling $0$ once. 

Using the expression (\ref{GF[Yj]}) for the generating function of the height in terms of the Bethe roots $Y_{j}$, we obtain the probability as a sum over eigenstates labelled by sets $J$ of $N$ integers between $1$ and $L$. For each set $J$, the change of variable $g=g_{[B,J],q}\to B$, whose Jacobian
\begin{equation}
\label{dg/dB}
\frac{B\,\partial_{B}g_{[B,J],q}}{g_{[B,J],q}}=\frac{L}{N}\,\langle1|(1-\mathbb{K}_{\p,q})^{-1}\,U_{\p,q}V_{\p,q}\rangle_{\p}
\end{equation}
is derived in \ref{appendix dB}, cancels a factor from the Gaudin determinant (\ref{det[int]}). The formulas of the previous section for various symmetric functions of the Bethe roots give our final result for stationary initial condition
\begin{eqnarray}
\label{P(Hi) stat}
&&\fl
\P_{\text{stat}}(H_{i}(t)=\tfrac{Ni}{L}+U)=\oint\frac{\rmd B}{2\rmi\pi B}\,\sum_{J}
\frac{(-1)^{\frac{N(N-1)}{2}}\prod_{j=0}^{N-1}(1-q^{j}/g_{\p,q})^{2}}{{{L}\choose{N}}}
\frac{\bar{\Pi}_{\p}^{i+1}\,V^{2}_{\p}}{\Pi^{*}_{\p}\,\Pi_{\p}^{N}\,g_{\p,q}^{U-N}}\\
&&\fl\hspace{5mm}
\times\rme^{t(1-q)\big(\frac{\eta_{\p}}{Y_{\p,q}'(1)}+\oint\frac{\rmd z}{2\rmi\pi z}\,A_{\p}(z)\,\big(\frac{1}{1-Y_{\p,q}(z)}-\frac{1}{1-qY_{\p,q}(z)}-\frac{1}{Y_{\p,q}'(1)}\,\frac{z}{1-z}\big)\big)+i\oint\frac{\rmd z}{2\rmi\pi z}\,A_{\p}(z)\,\log\big(\frac{1}{1-z}\,\frac{1-Y_{\p,q}(z)}{1-qY_{\p,q}(z)}\big)}\nonumber\\
&&\fl\hspace{5mm}
\times\frac{\rme^{-\oint\frac{\rmd z}{2\rmi\pi z}\,A_{\p}(z)\,\log\big(\frac{Y_{\p,q}(z)}{z}\big)+\frac{1}{2}\oint\frac{\rmd w}{2\rmi\pi w}\oint\frac{\rmd z}{2\rmi\pi z}\,A_{\p}(w)A_{\p}(z)\log\big(\frac{w z}{(w-z)^{2}}\,\frac{(Y_{\p,q}(w)-Y_{\p,q}(z))^{2}}{(Y_{\p,q}(w)-qY_{\p,q}(z))(Y_{\p,q}(z)-qY_{\p,q}(w))}\big)}}{\det(1-\mathbb{K}_{\p,q})}
\;,\nonumber
\end{eqnarray}
where $U\in\mathbb{Z}$, $\p=[B,J]$ and the summation is over all subsets $J$ of $\[1,L\]$ with $N$ elements. The coefficients $\Pi_{\p}$, $\bar{\Pi}_{\p}$, $\Pi^{*}_{\p}$, $V^{2}_{\p}$ defined in section~\ref{section TASEP} are meromorphic functions of $\p\in\R$. The function $Y_{\p,q}$ is the solution of (\ref{eq Y}), the fugacity $g_{\p,q}$ is defined in (\ref{g[Y]}), and the kernel of the integral operator $\mathbb{K}_{\p,q}$ is given in (\ref{K}). More generally, for a system starting at time $t=0$ with initial probabilities $P_{0}(\mathcal{C})$, one has
\begin{eqnarray}
\label{P(Hi) P0}
&&\fl
\P(H_{i}(t)=\tfrac{Ni}{L}+U)=\oint\frac{\rmd B}{2\rmi\pi B}\,\sum_{J}
\Theta_{\p,q}^{P_{0}}\,\Big(\prod_{j=0}^{N-1}(1-q^{j}/g_{\p,q})\Big)\\
&&\fl\hspace{3mm}
\times\frac{\bar{\Pi}_{\p}^{i}\,\rme^{t(1-q)\big(\frac{\eta_{\p}}{Y_{\p,q}'(1)}+\oint\frac{\rmd z}{2\rmi\pi z}\,A_{\p}(z)\,\big(\frac{1}{1-Y_{\p,q}(z)}-\frac{1}{1-qY_{\p,q}(z)}-\frac{1}{Y_{\p,q}'(1)}\,\frac{z}{1-z}\big)\big)+i\oint\frac{\rmd z}{2\rmi\pi z}\,A_{\p}(z)\,\log\big(\frac{1}{1-z}\,\frac{1-Y_{\p,q}(z)}{1-qY_{\p,q}(z)}\big)}}{\Pi_{\p}^{N}\,g_{\p,q}^{U-N}}\nonumber\\
&&\fl\hspace{3mm}
\times\frac{(1-q)^{N}\,\Pi_{\p}\bar{\Pi}_{\p}\,V^{2}_{\p}}{(-1)^{\frac{N(N-1)}{2}}\,\Pi^{*}_{\p}}\,\frac{\rme^{\frac{1}{2}\oint\frac{\rmd w}{2\rmi\pi w}\oint\frac{\rmd z}{2\rmi\pi z}\,A_{\p}(w)A_{\p}(z)\log\big(\frac{w z}{(w-z)^{2}}\,\frac{(Y_{\p,q}(w)-Y_{\p,q}(z))^{2}}{(Y_{\p,q}(w)-qY_{\p,q}(z))(Y_{\p,q}(z)-qY_{\p,q}(w))}\big)}}{\det(1-\mathbb{K}_{\p,q})}\;,\nonumber
\end{eqnarray}
where $\Theta_{\p,q}^{P_{0}}$ is defined by (\ref{Theta}).

The expression (\ref{P(Hi) stat}) was checked numerically up to order $q^{3}$ for all systems with $1\leq N<L\leq4$ and for all $i\in\[0,L\]$, $U\in\[-3,3\]$ with a generic value for $t$, against a numerical evaluation of (\ref{P[GF]}) with $\langle g^{H_{i}(t)}\rangle$ evaluated from (\ref{GF[M] P0}). The expression (\ref{P(Hi) P0}) was similarly checked numerically up to order $q^{2}$, for all possible initial states of the form $|P_{0}\rangle=|\mathcal{C}_{0}\rangle$, $\mathcal{C}_{0}\in\Omega$.

As shown in section~\ref{section Y}, $Y_{\p,q}$ is at each order in $q$ a polynomial in the $\alpha_{\p,m}$, $m\in\Z^{*}$ with coefficients independent of the point $\p\in\R$. This is then also the case for all the contour integrals involving $A_{\p}$ and for the Fredholm determinant $\det(1-\mathbb{K}_{\p,q})$ in (\ref{P(Hi) P0}). Writing for short (\ref{P(Hi) P0}) as $\oint\frac{\rmd B}{2\rmi\pi B}\,\sum_{J}Z_{\p}$ and recalling the results of section~\ref{section TASEP}, we observe that $Z_{\p}$ is, at each order in $q$, a meromorphic function of $\p\in\R$ with poles $[0,J]$, $[B_{*},J]$, plus essential singularities at points $\p=[\infty,J]$ coming from the factor $\exp(t(1-q)\frac{\eta_{\p}}{Y_{\p,q}'(1)})$. Furthermore, as mentioned at the end of section~\ref{section TASEP}, while the function $\p\mapsto V^{2}_{\p}/\Pi^{*}_{\p}$ has poles at some points $\p=[B_{*},J]$, the differential $\rmd B\,V^{2}_{\p}/\Pi^{*}_{\p}$ with $\p=[B,J]$ is however holomorphic at those points. Thus, we conclude that the meromorphic differential $Z_{\p}\,\rmd B$ only has poles (and essential singularities) at points $[0,J]$ and $[\infty,J]$, at least at each order in $q$.

Summing over $J$, the differential $\sum_{J}Z_{[B,J]}\,\rmd B$ in (\ref{P(Hi) P0}) then has a (multiple) pole at $B=0$, an essential singularity at $B=\infty$, and no other singularity. The contour for $B$ can thus be moved freely in (\ref{P(Hi) P0}) provided $0$ stays inside of the contour. This was already the case for TASEP \cite{P2020.2}. We note that the absence of other poles in (\ref{P(Hi) P0}) depends crucially on our choice for the fugacity $\rme^{\gamma}=g_{\p,q}$ with $g_{\p,q}$ defined in (\ref{g[Y]}). Indeed, any change of variable $B\to C$ with $C=B+\mathcal{O}(q)$ would give terms $\mu_{[C,J]}$, introducing an additional pole at $C=B_{*}$ for the differential $\sum_{J}Z_{[B,J]}\,\rmd B$, since e.g. $B\partial_{B}\Pi_{[B,J]}=\Pi_{[B,J]}(1+\frac{L}{L-N}\,\mu_{[B,J]})$. Additionally, the cancellation by the Jacobian (\ref{dg/dB}) of the factor $\langle1|(1-\mathbb{K}_{\p,q})^{-1}\,U_{\p,q}V_{\p,q}\rangle_{\p}$ coming from the Gaudin determinant (\ref{det[int]}), and whose expansion in powers of $q$ contains coefficients $\mu_{\p}$, is crucial for the absence of extra poles at $\p=[B_{*},J]$.

\noindent{\bf Remark 1:} The expression (\ref{P(Hi) P0}) for the probability, $\oint\frac{\rmd B}{2\rmi\pi B}\,\sum_{J}Z_{p}$ for short, with an integration over a closed curve in $\C$ encircling $0$, can alternatively be written as $\oint_{\gamma}\frac{\rmd B}{2\rmi\pi B}\,Z_{\p}$, where the contour $\gamma\subset\R$ is now a reunion of simple closed curves splitting each connected component of the Riemann surface $\R$ into a domain containing all the points $[0,J]$ (with appropriate identifications at branch points) and another domain containing all the points $[\infty,J]$ of the connected component. Indeed, lifting a closed curve around $0$ to the sheet $\C_{J}$ of $\R$ by the inverse of the covering map $[B,J]\mapsto B$ produces a path $\gamma_{J}$ on $\R$. Considering then a point $\p_{0}\in\R$ sent to $0$ by $[B,J]\mapsto B$, and the identification set $S=\{J,\p_{0}=[0,J]\}$, the curve $\bigcup_{J\in S}\gamma_{J}$ is a closed contour on $\R$ encircling $\p_{0}$. Finally, the union for all possible $J$ corresponding to a given connected component of $\R$ of the paths $\gamma_{J}$ produces closed curves around all the points sent to $0$ by the covering map, which can be deformed into a single closed curve, which must however not cross the points $[\infty,J]$ because of the singularity of $Z_{\p}\rmd B$ at those points.

\noindent{\bf Remark 2:} For TASEP, the extra factor $\prod_{j=0}^{N-1}(1-q^{j}/g_{\p,q})=1-1/g_{\p,0}$ in (\ref{P(Hi) P0}) can be removed by considering instead the cumulative probability $\P(H_{i}(t)\geq\tfrac{Ni}{L}+U)$. For arbitrary $q$, using the $q$-binomial theorem $\prod_{j=0}^{N-1}(1-q^{j}/g)=\sum_{k=0}^{N}q^{\frac{k(k-1)}{2}}(-1)^{k}g^{-k}{{N}\choose{k}}_{q}$ and the identity $\sum_{k=0}^{j}q^{\frac{k(k-1)}{2}}(-1)^{k}g^{-k}{{N}\choose{k}}_{q}{{j-k+N-1}\choose{N-1}}_{q}=\delta_{j,0}$, we note that the factor $\prod_{j=0}^{N-1}(1-q^{j}/g_{\p,q})$ can similarly be removed from (\ref{P(Hi) P0}) at the price of considering instead the generating function $\langle{{H_{i}(t)-\frac{Ni}{L}-U+N-1}\choose{N-1}}_{q}\rangle$.
\end{subsection}
\end{section}

\begin{section}{Conclusion}
In this paper, we have extended to ASEP the Riemann surface approach introduced in \cite{P2020.2} for TASEP. We obtained an expression (\ref{P(Hi) P0}) for the probability of the ASEP height function as a contour integral of a meromorphic differential on the Riemann surface $\R$ characterizing the integrability of TASEP. The solution involves a function $Y_{\p,q}$ mapping Bethe roots of TASEP to Bethe roots of ASEP, which is solution of the integro-differential equation (\ref{eq Y}). The whole construction is based on an all order perturbative expansion in the hopping rate $q$.

Due to the perturbative nature of the solution, it is not clear whether the expressions obtained in this paper are suitable for taking the large $L,N$ limits, especially within the scaling $1-q\sim1/\sqrt{L}$ leading to the KPZ equation, which interpolates between the equilibrium fixed point and the KPZ fixed point. Additionally, the expressions obtained are very singular when the parameter $B$ goes to zero because of the coefficients $\alpha_{\p,m}$, $\p=[B,J]$ appearing in $Y_{\p,q}$. This prevents to recover in an efficient way known expressions \cite{DM1997.1,P2008.1,P2010.1} for stationary large deviations, which correspond for TASEP to the point $B=0$ of the sheet $J=\[1,N\]$ of $\R$, and for which functional equations somewhat analogous to (\ref{eq Y}) have been found \cite{P2010.1,LM2011.1}. A non-perturbative solution may be needed to resolve these issues, with the construction of a well defined compact Riemann surface distinct from $\R$ and characterizing the Bethe equations of ASEP for arbitrary $q$, even in the regions $B\to0$ of the phase space. It might be necessary for this to consider also the Bethe equations ``beyond the equator'' \cite{PS1999.1}, which were used to compute stationary large deviations of the current for ASEP in \cite{P2010.1}.
\end{section}

\appendix
\begin{section}{Derivatives with respect to \texorpdfstring{$B$}{B}}
\label{appendix dB}
In this section, we compute $\partial_{B}Y_{\p,q}$ with $\p=[B,J]$ and show that the Jacobian $\partial_{B}g_{\p,q}$ is given by (\ref{dg/dB}).

\begin{subsection}{\texorpdfstring{$\partial_{B}Y_{\p,q}$}{dY/dB}}
We set $\p=[B,J]$ everywhere in this section, and compute the partial derivative of $Y_{\p,q}(y)$ with respect to the parameter $B$. For simplicity of the derivation, it is convenient to start with (\ref{eq g Y}) instead of the closed equation (\ref{eq Y}) for $Y_{\p,q}$, and to use the variable $Z=Y_{\p,q}(y)$ instead of $y$ (since $Y_{\p,q}(y)=y+\mathcal{O}(q)$, the inverse function $Y_{\p,q}^{-1}$ is well defined at any order in $q$). Since $A_{\p}$ depends on $B$, we use first (\ref{sum[int] regular}) to express the contour integral as a sum. One has
\begin{equation}
\frac{g_{\p,q}}{B}
=\frac{(1-Y_{\p,q}^{-1}(Z))^{L}}{Y_{\p,q}^{-1}(Z)^{N}}\,\Big(\frac{1-qZ}{1-Z}\Big)^{L}\prod_{k\in J}\frac{Z-q\,Y_{\p,q}(y_{k}(B))}{Y_{\p,q}(y_{k}(B))-q\,Z}\;.
\end{equation}
Taking the logarithmic derivative with respect to $B$ and setting $Z=Y_{\p,q}(y)$ afterwards leads to
\begin{eqnarray}
\label{dBlog(g)}
&&\fl\hspace{2mm}
\partial_{B}\log\Big(\frac{g_{\p,q}}{B}\Big)
=-\frac{N+(L-N)y}{y(1-y)}\,(\partial_{B}Y_{\p,q}^{-1})(Y_{\p,q}(y))\\
&&\fl\hspace{23mm}
-\sum_{k\in J}\Big(\frac{q}{Y_{\p,q}(y)-q\,Y_{\p,q}(y_{k}(B))}+\frac{1}{Y_{\p,q}(y_{k}(B))-qY_{\p,q}(y)}\Big)\,\partial_{B}(Y_{\p,q}(y_{k}(B)))\;.\nonumber
\end{eqnarray}
Subtracting from the equation above the same equation at $y=-\frac{N}{L-N}$ eliminates the term $\partial_{B}\log\big(\frac{g_{\p,q}}{B}\big)$. Using
\begin{equation}
\label{dBY^-1}
(\partial_{B}Y_{\p,q}^{-1})(Y_{\p,q}(y))=-\frac{\partial_{B}Y_{\p,q}(y)}{Y_{\p,q}'(y)}\;,
\end{equation}
one has
\begin{equation}
\frac{\partial_{B}Y_{\p,q}(y)}{U_{\p,q}(y)}=\sum_{k\in J}\frac{K_{\p,q}^{\dagger}(y,y_{k}(B))}{U_{\p,q}(y_{k}(B))}\,\partial_{B}(Y_{\p,q}(y_{k}(B)))\;,
\end{equation}
where $U_{\p,q}$ is defined in (\ref{U}) and $K_{\p,q}^{\dagger}(y,z)=K_{\p,q}(z,y)$ with $K_{\p,q}$ defined in (\ref{K}). Using furthermore
\begin{equation}
\label{dBY(yk(B))}
\partial_{B}(Y_{\p,q}(y_{k}(B)))=(\partial_{B}Y_{\p,q})(y_{k}(B))+Y_{\p,q}'(y_{k}(B))\partial_{B}y_{k}(B)
\end{equation}
and
\begin{equation}
\label{yk'(B)}
B\partial_{B}y_{k}(B)=\frac{y_{k}(B)(1-y_{k}(B))}{N+(L-N)y_{k}(B)}\;,
\end{equation}
we finally obtain
\begin{equation}
\label{eq BdBY sum}
\fl\hspace{5mm}
\frac{B\partial_{B}Y_{\p,q}(y)}{U_{\p,q}(y)}-\sum_{k\in J}K_{\p,q}^{\dagger}(y,y_{k}(B))\,\frac{(B\partial_{B}Y_{\p,q})(y_{k}(B))}{U_{\p,q}(y_{k}(B))}
=\sum_{k\in J}K_{\p,q}^{\dagger}(y,y_{k}(B))\;.
\end{equation}
Defining the operator $\mathbb{K}_{\p,q}^{\dagger}$, adjoint of the operator $\mathbb{K}$ defined in (\ref{K op}), and acting on meromorphic functions by
\begin{equation}
(\mathbb{K}_{\p,q}^{\dagger}f)(y)=\sum_{k\in J}K_{\p,q}^{\dagger}(y,y_{k}(B))f(y_{k}(B))\;,
\end{equation}
equation (\ref{eq BdBY sum}) rewrites more compactly as
\begin{equation}
(1-\mathbb{K}_{\p,q}^{\dagger})\frac{B\partial_{B}Y_{\p,q}}{U_{\p,q}}=\mathbb{K}_{\p,q}^{\dagger}\,1\;.
\end{equation}
Since $K_{\p,q}^{\dagger}(y,z)=\mathcal{O}(q)$, the operator $1-\mathbb{K}_{\p,q}^{\dagger}$ can be inverted at least perturbatively in $q$, and we obtain
\begin{equation}
\label{BdBY}
\frac{B\partial_{B}Y_{\p,q}}{U_{\p,q}}=(1-\mathbb{K}_{\p,q}^{\dagger})^{-1}\mathbb{K}_{\p,q}^{\dagger}\,1
=\sum_{r=1}^{\infty}(\mathbb{K}_{\p,q}^{\dagger})^{r}\,1\;.
\end{equation}
Equivalently, using (\ref{dBY(yk(B))}) and (\ref{yk'(B)}), one has
\begin{equation}
\label{BdBY k}
\fl\hspace{10mm}
\frac{B\partial_{B}(Y_{\p,q}(y_{k}(B)))}{U_{\p,q}(y_{k}(B))}=\Big((1-\mathbb{K}_{\p,q}^{\dagger})^{-1}\,1\Big)(y_{k}(B))
=\Bigg(\sum_{r=0}^{\infty}(\mathbb{K}_{\p,q}^{\dagger})^{r}\,1\Bigg)(y_{k}(B))\;.
\end{equation}
\end{subsection}

\begin{subsection}{\texorpdfstring{$\partial_{B}g_{\p,q}$}{dg/dB}}
We now show that the Jacobian $\partial_{B}g_{\p,q}$ with $\p=[B,J]$ is indeed given by (\ref{dg/dB}). Starting with (\ref{dBlog(g)}) and using (\ref{dBY^-1}), one has
\begin{equation}
\fl\hspace{15mm}
\frac{B\partial_{B}g_{\p,q}}{g_{\p,q}}
=1+\frac{B\partial_{B}Y_{\p,q}(y)}{U_{\p,q}(y)}
-\sum_{k\in J}X_{\p,q}(y_{k}(B),y)\,B\partial_{B}(Y_{\p,q}(y_{k}(B)))\;,
\end{equation}
where $X_{\p,q}$ is defined in (\ref{X}). Then, using (\ref{BdBY}),  (\ref{BdBY k}), we obtain
\begin{eqnarray}
&&
\frac{B\partial_{B}g_{\p,q}}{g_{\p,q}}
=((1-\mathbb{K}_{\p,q}^{\dagger})^{-1}\,1)(y)\\
&&\hspace{23mm}
-\sum_{k\in J}U_{\p,q}(y_{k}(B))\,X_{\p,q}(y_{k}(B),y)\,\Big((1-\mathbb{K}_{\p,q}^{\dagger})^{-1}\,1\Big)(y_{k}(B))\nonumber
\end{eqnarray}
for arbitrary $y$. Setting $y=y_{j}(B)$, summing over $j\in J$ and using (\ref{eq Y UVX}) leads to
\begin{eqnarray}
&&\fl\hspace{5mm}
N\,\frac{B\partial_{B}g_{\p,q}}{g_{\p,q}}
=\sum_{j\in J}((1-\mathbb{K}_{\p,q}^{\dagger})^{-1}\,1)(y_{j}(B))\\
&&\hspace{5mm}
-\sum_{k\in J}\Big(1-L\,U_{\p,q}(y_{k}(B))\,V_{\p,q}(y_{k}(B))\Big)\,\Big((1-\mathbb{K}_{\p,q}^{\dagger})^{-1}\,1\Big)(y_{k}(B))\;,\nonumber
\end{eqnarray}
which finally implies (\ref{dg/dB}).
\end{subsection}
\end{section}


\end{document}